\documentclass[preprints,article,submit,pdftex,moreauthors]{Definitions/mdpi} 
\firstpage{1} 
\pubvolume{1}
\issuenum{1}
\articlenumber{0}
\pubyear{2025}
\copyrightyear{2025}
\datereceived{ } 
\daterevised{ } % Comment out if no revised date
\dateaccepted{ } 
\datepublished{ } 
\hreflink{https://doi.org/} % If needed use \linebreak
\usepackage{makecell}    % For multi-line cells
\usepackage{adjustbox}   % For scaling table to fit page
\usepackage{caption}     % For caption formatting (optional but recommended)

\usepackage{lipsum}
\usepackage{amsmath, amssymb, amsfonts}

\usepackage{listings}
\lstdefinestyle{mypython}{
  language=Python,
  basicstyle=\ttfamily\small,
  keywordstyle=\color{blue},
  stringstyle=\color{red},
  commentstyle=\color{gray},
  numbers=left,
  numberstyle=\tiny\color{gray},
  stepnumber=1,
  numbersep=5pt,
  tabsize=4,
  showstringspaces=false,
  breaklines=true,
  frame=single,
  keywords={yield}
}
\usepackage{algorithm}
\usepackage{algpseudocode}

\usepackage{graphicx}
\usepackage{caption}
\usepackage{subcaption}

\usepackage[simplified]{pgf-umlcd}
\usepackage{tikzsymbols}
\usetikzlibrary{
  arrows, positioning, patterns, arrows.meta,
  calc, shadows,
  shapes.geometric, shapes.symbols, shapes.arrows,
  shapes.multipart, shapes.callouts, shapes.misc
}

\newcommand{\ignore}[1]{{}}

\usepackage{float}  % allows [H] option to force float placement

\usepackage{pgfplots}
\pgfplotsset{compat=1.18}
\usepackage{adjustbox}

\usepackage{colortbl}
\definecolor{lightgreen}{RGB}{200,255,200}
\definecolor{lightred}{RGB}{255,200,200}

\usepackage{siunitx}
\usetikzlibrary{angles,quotes}

\usepgfplotslibrary{groupplots}
\pgfplotsset{compat=1.18}

\usepackage[utf8]{inputenc}
\usepackage{pgfplots}
\pgfplotsset{compat=1.18} % PGFPlots compatibility
\usepackage{amsmath} % Required for math environments, though not strictly used here

\usepackage{booktabs}
\renewcommand{\arraystretch}{1.3}

\Title{Black-Box Bug-Amplification for Multithreaded Software}

\TitleCitation{Black-Box Bug-Amplification for Multithreaded Software}

\Author{
Yeshayahu~Weiss$^{1\dagger*}$\orcidA{}, 
Gal~Amram$^{2\ddagger} $\orcidE{}, 
Achiya~Elyasaf$^{3\ddagger}$\orcidF{}
Eitan~Farchi$^{2\ddagger} $\orcidD{}, 
Oded~Margalit$^{1\ddagger} $\orcidC{}, 
Gera~Weiss$^{1\ddagger}$\orcidB{}, 
}

\AuthorNames{Yeshayahu Weiss, Gal Amram, Achiya Elyasaf, Eitan Farchi, Oded Margalit, Gera Weiss}

\AuthorCitation{Weiss, Y.; Amram, G.; Elyasaf, A.; Farchi, E.; Margalit, O.; Weiss, G.}

\address{%
$^{1}$ \quad Department of Computer Science, Ben-Gurion University of the Negev, Be'er Sheva, Israel; \texttt{weissye@post.bgu.ac.il};  \texttt{geraw@bgu.ac.il}; \texttt{odedm@bgu.ac.il}; \\
$^{2}$ \quad IBM Research, Haifa, Israel; \texttt{farchi@il.ibm.com}; 
\texttt{gal.amram@ibm.com} \\
$^{3}$ \quad Department of Software and Information Systems Engineering, Ben-Gurion University of the Negev, Be'er Sheva, Israel; \texttt{achiya@bgu.ac.il} 
}

\corres{Correspondence: \texttt{weissye@post.bgu.ac.il}}

\firstnote{Current address: Ben-Gurion University of the Negev, Be'er Sheva, Israel.}
\secondnote{These authors contributed equally to this work.}

\abstract{
Bugs, especially those in concurrent systems, are often hard to reproduce because they manifest only under rare conditions. Testers frequently encounter failures that occur only under specific inputs, even when occurring with low probability. We propose an approach to systematically amplify the occurrence of such elusive bugs. We treat the system under test as a black-box and use repeated trial executions to train a predictive model that estimates the probability of a given input configuration triggering a bug. We evaluate this approach on a dataset of 17 representative concurrency bugs spanning diverse categories. Several model-based search techniques are compared against a brute-force random sampling baseline. Our results show that an ensemble of regression models can significantly increase bug occurrence rates across nearly all scenarios, often achieving an order-of-magnitude improvement over random sampling. The contributions of this work include: (i) a novel formulation of bug-amplification as a rare-event regression problem; (ii) an empirical evaluation of multiple techniques for amplifying bug occurrence, demonstrating the effectiveness of model-guided search; and (iii) a practical, non-invasive testing framework that helps practitioners expose hidden concurrency faults without altering the internal system architecture.
}


\begin{document}

\section{Introduction}
\label{sec:introduction}

Bugs that manifest nondeterministically, sometimes referred to as \emph{Heisenbugs}~\cite{Gray1985} or intermittent bugs \cite{intermitent}, pose a significant challenge for debugging and validation in complex software systems. This difficulty is particularly pronounced for \emph{concurrency bugs}, which typically arise only under rare thread interleavings or delicate timing conditions. In practice, developers rely on techniques such as manual code inspection and brute-force stress testing to uncover such failures. Although stress testing may occasionally expose these elusive faults, it offers no guarantees of detecting and often fails to detect bugs that appear only under constrained conditions. As a consequence, critical concurrency issues can remain unresolved for extended periods, undermining confidence in the system’s reliability.

In this paper, we \emph{maximize the empirical failure probability} observed during testing, under a fixed execution budget. We refer to this goal as \emph{bug amplification}. In contrast to approaches such as rare-event simulation~\cite{rare-event-simulation} and statistical model
checking~\cite{statistical-model-checking} which 
typically rely on internal instrumentation, white-box knowledge, or formal specifications; our method operates in a fully black-box manner. We assume no access to source code or internal system behavior. Instead, we systematically vary input parameters, such as workload configurations and timing-related settings, to increase the likelihood that a latent bug will manifest during execution.

Despite progress in the field, reliably exposing concurrency bugs in real-world systems remains an open challenge~\cite{Kumar2025}. Systematic concurrency testing tools attempt to exhaustively explore possible thread schedules and can enable deterministic replay of bugs once discovered. However, their scalability is limited by the combinatorial explosion of scheduling interleavings. Alternatively, randomized scheduling introduces noise to execution timing and has shown improved coverage compared to naive stress tests~\cite{Burckhardt2010, zhao2025selectively}, yet remains fundamentally probabilistic and may still miss deeply hidden bugs. Record-and-replay tools log nondeterministic events during execution for later reproduction, but their performance overhead and requirement for tightly controlled environments make them impractical in many settings. Collectively, these approaches fall short of providing a general, scalable solution for reliably triggering elusive concurrency failures.

To address this gap, we introduce a novel approach that frames bug amplification as a black-box optimization problem over the system’s input space. Rather than modifying internal code or 
instrumenting it, we run the system repeatedly under different input configurations and observe whether a failure occurs. These outcomes are used to train a predictive model estimating the probability of failure as a function of the input parameters. The model then guides the generation of future test inputs, focusing resources on regions of the input space more likely to expose the bug.

Casting this task as a regression problem presents unique difficulties. The target function, a binary indicator of failure, is one with an extremely sparse positive signal, often yielding zero in most regions of the input space. Even for failure-prone configurations, the bug may only appear with low probability due to nondeterministic execution. To cope with this challenge, we perform multiple trials for each input and use the average failure rate as a noisy estimate of its true failure probability. This allows us to apply regression algorithms despite the underlying stochasticity and imbalance, though it necessitates robust modeling techniques capable of tolerating label noise and extreme skew.

To evaluate the proposed strategy, we curated a benchmark of 17 concurrency bugs spanning a comprehensive taxonomy of bug symptoms and their underlying causes. These bugs, drawn from real-world and synthetic sources, cover a variety of symptoms (e.g., deadlocks, crashes, data races) and underlying causes (e.g., incorrect synchronization, ordering violations). For each problem, we identified key input parameters that influence bug manifestation and tuned the system so that failures occur with low probability under default settings. This controlled setup enables rigorous assessment of amplification techniques under realistic yet challenging conditions.

We applied several model-based search techniques to the benchmark, including linear regression, decision trees, and nonlinear ensemble methods, and compared them against a baseline of brute-force random sampling. Under identical budget constraints, a stacked ensemble of classifiers consistently achieved the best overall performance, substantially increasing bug manifestation rates across the majority of scenarios.

\noindent \textbf{This work makes the following contributions:}
\begin{itemize}
    \item \textbf{Benchmark and Problem Formulation:} We introduce a curated dataset of 17 concurrency bugs and frame the failure-triggering task as a regression problem with sparse positives and stochastic labels---posing distinct challenges for conventional learners.
    \item \textbf{Evaluation of Amplification Techniques:} We systematically compare several model-guided search strategies and show that ensemble-based learning significantly improves bug-triggering probability within practical testing budgets.
    \item \textbf{Practical, Black-box Testing Framework:} Our approach treats the system under test as a black box, requiring no code changes or instrumentation, making it readily applicable in real-world testing workflows.
\end{itemize}

The remainder of this paper is organized as follows.
\autoref{sec:state-of-the-art} reviews the state of the art in bug reproduction, presenting leading techniques and key challenges in the field.
\autoref{sec:ontologies} provides a detailed classification of concurrency bug types relevant to our study.
\autoref{sec:benchmark} summarizes the benchmark problems used in our evaluation, outlining the criteria for selection.
\autoref{sec:simulate} describes the core research methods, focusing on the modeling of interleaving in multithreaded code.
\autoref{sec:tested-approaches} introduces the four bug-amplification methods that we developed and applied, and provides implementation and configuration specifics.
\autoref{sec:results} presents the experimental results and discusses their implications.
Finally, the paper concludes with a summary of findings, a detailed list of the limitations of our approach, and directions for future research.

%TBC - Eitan to continue reading from here

\section{State of the Art in Bug Reproduction}
\label{sec:state-of-the-art}

Reproducing nondeterministic concurrency failures remains a central challenge in software testing. These bugs typically occur only under rare thread interleavings or specific combinations of environmental and input parameters, making them elusive and difficult to diagnose~\cite{ramesh2025unveiling}. 

Traditional techniques such as stress testing, heuristic scheduling perturbations, and detailed logging have been widely used in practice, but they offer no guarantees and are often insufficient for reliably exposing such rare failures~\cite{Godefroid2015}. CARDSHARK~\cite{Han2024}, for example, demonstrates how even kernel-level bugs may remain unstable without explicit noise control or scheduling alignment.

\textbf{Industry Practice.} When developers encounter rare failures in production, a common response is to attempt reproduction via repeated testing under varied conditions, manipulating input sizes, concurrency levels, or hardware settings~\cite{Bianchi2017ConCrashSearch}. Logging may provide diagnostic clues, but even lightweight instrumentation, such as coverage or profiling hooks, can perturb timing behavior enough to mask or induce concurrency bug manifestation~\cite{Rasheed2023}. Kernel-level concurrency testing frameworks such as the eBPF-based technique by Xu et al.~\cite{Xu2025} offer promising lightweight instrumentation for observing concurrency bugs in real-world deployments.

\textbf{Systematic Exploration.} Research tools such as CHESS~\cite{Musuvathi2008}, Nekara~\cite{Shashidhar2021}, and Fray~\cite{Kumar2025} aim to improve bug reproducibility by exhaustively exploring thread schedules in bounded spaces. CHESS is a pioneering systematic testing tool for multithreaded Windows applications that explores all interleavings under a given bound. Nekara is an open-source, cross-platform library (2021) that enables developers to define semantics for concurrency primitives and systematically explore schedules in a controlled, repeatable manner. Fray, introduced in 2025, offers efficient black-box schedule control and instrumentation for JVM-based systems. These tools can replay discovered interleavings deterministically, a key advantage for debugging, but they require either source or binary instrumentation and do not scale well with large programs or vast input spaces.

\textbf{Probabilistic Scheduling and Sampling.} Techniques like Probabilistic Concurrency Testing (PCT)~\cite{Lee2022}, iterative schedule fuzzing~\cite{Elmas2013}, and directional scheduling of synchronization primitives in Go programs~\cite{Chen2023} attempt to bias execution toward schedules more likely to reveal bugs. While these methods can improve exposure rates, they remain largely unguided by feedback from prior executions.

\textbf{Learning-Based Approaches.} Recent advances have begun exploring machine learning for bug localization and input generation~\cite{Li2019DeepFL, Bottinger2018LearnFuzz}, but most treat the system as a white box or focus on symbolic execution, mutation, or coverage estimation~\cite{Amalfitano2023}. By contrast, our method treats the system as a black-box and explicitly aims to maximize the empirical failure probability via predictive models over the input domain.

Building on these advances, our work treats bug reproduction as a noisy optimization problem over inputs, training predictive classifiers to guide search. Instead of exploring schedules, we vary inputs and use learned models to amplify bug occurrence rates within constrained testing budgets, improving reproducibility and efficiency.

\section{Types of Concurrency Bugs}    
\label{sec:ontologies}

Following prior work such as~\cite{10.1145/2954679.2872374}, we introduce a taxonomy to support the evaluation of our bug-amplification techniques. This taxonomy classifies concurrency bugs along two orthogonal dimensions: \textit{observable effect} and \textit{root cause}.

%This ontology provides a conceptual framework for organizing and reasoning about the diverse ways in which concurrency can go wrong.
In detail, the observable effect axis captures how a concurrency bug manifests at runtime, i.e., the observable effect or symptom from the system's perspective. The second or root cause axis reflects the underlying cause of the failure, identifying the specific logic error or design flaw in the program's synchronization or concurrency control.  

Classification of the observable effects of the concurrent bug is done using the following categories.

%\subsection*{Classification by Observable Effect}
\begin{itemize}
    \item \textbf{Deadlock:} A system state in which two or more threads are indefinitely blocked, each waiting for a resource that will never become available, e.g., because it is held by another. The system halts and cannot make further progress.

    \item \textbf{Unexpected Data:} Shared variables take on incorrect or inconsistent values due to unsynchronized access, race conditions, or improper interleaving of reads and writes.

    \item \textbf{Concurrent Access:} Multiple threads enter a critical section simultaneously, violating mutual exclusion and potentially corrupting shared state or breaking invariants.
\end{itemize}

%TBC - Eitan continue reading 
\noindent
While the classification of the root cause of the concurrent bug is done using the following categories. 
%\subsection*{Classification by Root Cause}

\begin{itemize}
    \item \textbf{Missing or Weak Guarding:} Inadequate protection of critical sections, often due to absent atomicity checks, incorrect condition synchronization, or overreliance on scheduling assumptions.

    \item \textbf{Non-Atomic Operations on Shared State:} Access to shared data is implemented via sequences of non-atomic operations, allowing interleaving by other threads to interfere with correctness.

    \item \textbf{Incorrect Command Ordering:} Synchronization operations are issued in the wrong order, violating required temporal constraints. For example, a thread signals a condition before another begins waiting for it.

    \item \textbf{Misuse of Concurrency Primitives:} synchronization constructs such as locks, semaphores, and condition variables are used incorrectly, e.g., in unintended contexts, or in ways that violate their semantics.
\end{itemize}

The cross-product of these two axes yields twelve distinct categories of concurrency bugs, each representing a unique pairing of effect and cause. Table~\ref{tab:concurrency-classification} summarizes the distribution of our benchmark problems across this taxonomy, with each problem assigned to the cell corresponding to its observed effect and inferred root cause. As a root cause may have more than a single effect, a problem index may appear twice in the same column, but not in the same row.

\begin{table}[ht]
\centering
\caption{Classification of concurrency problems by \textit{Effect} (rows) and \textit{Root Cause} (columns), showing the problem number and name. Note that some problems may produce multiple effects (e.g., Problems 12 and 4).}
\begin{adjustbox}{max width=\textwidth}
\begin{tabular}{|l|p{3.8cm}|p{3.4cm}|p{4.2cm}|p{3.8cm}|}
\hline
\rowcolor{gray!25}
\multicolumn{1}{|c|}{\textbf{Effect $\backslash$ Root Cause}} & 
\begin{tabular}{@{}c@{}}\textbf{Missing/Weak Guard}\end{tabular} & 
\textbf{Non-Atomic Op.} & 
\textbf{Incorrect Ordering} & 
\textbf{Misuse of Primitives} \\
\hline
\hline
\textbf{Deadlock} &
\begin{tabular}{@{}l@{}}6 (If-Not-While)\\8 (Lost Signal)\\17 (Sleeping Guard)\end{tabular} &
\begin{tabular}{@{}l@{}}11 (Race-To-Wait)\end{tabular} &
\begin{tabular}{@{}l@{}}7 (Lock Order Inversion)\\16 (Signal-Then-Wait)\end{tabular} &
\begin{tabular}{@{}l@{}}2 (Broken Barrier)\\5 (Flagged Deadlock)\end{tabular} \\
\hline
\textbf{Unexpected Data} &
\begin{tabular}{@{}l@{}}6 (If-Not-While)\\9 (Partial Lock)\end{tabular} &
\begin{tabular}{@{}l@{}}12 (Racy Increment)\\14 (Shared Counter)\end{tabular} &
\begin{tabular}{@{}l@{}}4 (Delayed Write)\end{tabular} &
\begin{tabular}{@{}l@{}}1 (Atomicity Bypass)\end{tabular} \\
\hline
\textbf{Concurrent Access} &
\begin{tabular}{@{}l@{}}3 (Broken Peterson)\\15 (Shared Flag)\end{tabular} &
\begin{tabular}{@{}l@{}}12 (Racy Increment)\\14 (Shared Counter)\end{tabular} &
\begin{tabular}{@{}l@{}}4 (Delayed Write)\end{tabular} &
\begin{tabular}{@{}l@{}}10 (Phantom Permit)\\13 (Semaphore Leak)\end{tabular} \\
\hline
\end{tabular}
\end{adjustbox}
\label{tab:concurrency-classification}
\end{table}

The inclusion of at least one benchmark problem in each of the twelve cells of the classification matrix ensures that our taxonomy is comprehensively represented. This guarantees that the analysis spans all combinations of observable effects and root causes, ensuring broad and representative coverage of concurrency failure modes.

%%%%%%%%%%%%%%%%%%%%%%%%%%%%%%%%%%%%%%%%%%%%%%%%%%%%%%%%%%%%%
\section{Summary of the Benchmark Problems}     
\label{sec:benchmark}

To evaluate our ability to amplify and detect failure cases in multithreaded systems, we assembled a benchmark that spans the primary classes of concurrency faults.  Each problem instance illustrates a distinct error pattern, and the accompanying description clarifies the type of defect it represents. The benchmark is available in a \textit{GitHub repository}\footnote{\url{https://github.com/geraw/bug_amp}} 

The benchmark is based on the canonical puzzles from \textit{The~Deadlock~Empire}\footnote{\url{https://deadlockempire.github.io/\#menu}}, an interactive collection of multithreading challenges that can be executed step-by-step.  To achieve the broader coverage outlined in the previous section, we extended this initial set with additional cases gathered from the literature and custom-crafted variants, until all combinations of Effect (Deadlock, Unexpected Data, Concurrent Access) and Root Cause (Missing or Weak Guarding, Non-Atomic Operations, Incorrect Command Ordering, Misuse of Concurrency Primitives) were represented.  

\autoref{sec:benchmark-detailed} provides a full description of each of the 17 concurrency problems enumerated below.
 For every problem, we explicitly document (i) the scenario, (ii) its observable effect, (iii) the underlying root cause according to our taxonomy, and (iv) a concise insight that summarizes the key lesson.  This curated collection provides a balanced testbed for assessing failure-amplification techniques across the full spectrum of concurrency bugs.

\begin{description}

\item[Atomicity Bypass:]
A thread releases a lock before completing a read-modify-write, leading to data corruption despite apparent locking. See Section~\ref{sec:prob17}.

\item[Broken Barrier:]
Improper barrier reuse or reset causes some threads to wait forever, expecting others to arrive. See Section~\ref{sec:prob6}.

\item[Broken Peterson:]
Incorrect implementation of Peterson's algorithm allows both threads to enter the critical section. See Section~\ref{sec:prob11}.

\item[Delayed Write:]
Operations are reordered due to compiler or logic flaws, leading to stale reads or broken invariants. See Section~\ref{sec:prob10}.

\item[Flagged Deadlock:]
Threads use flags and spin loops incorrectly, creating interleaving paths that deadlock. See Section~\ref{sec:prob5}.

\item[If-Not-While:]
A thread waits using an \texttt{if} condition instead of a \texttt{while} loop, leading to missed signals and unsafe access. See Section~\ref{sec:prob8}.

\item[Lock Order Inversion:]
Classic deadlock: threads acquire two locks in opposite order, causing circular wait. See Section~\ref{sec:prob1}.

\item[Lost Signal:]
A thread sends a signal before another begins waiting on a condition variable; the signal is lost, causing a deadlock. See Section~\ref{sec:prob12}.

\item[Partial Lock:]
Only part of the critical section is protected by a lock; race conditions still occur. See Section~\ref{sec:prob4}.

\item[Phantom Permit:]
A semaphore is released without a corresponding \texttt{Wait}, allowing more threads than expected to enter the critical section. See Section~\ref{sec:prob16}.

\item[Race-To-Wait:]
Threads race to increment a shared counter and both wait on a condition that never becomes true due to non-atomic updates. See Section~\ref{sec:prob13}.

\item[Shared Flag:]
A single boolean flag is used for synchronization without proper mutual exclusion, allowing concurrent access. See Section~\ref{sec:prob3}

\item[Signal-Then-Wait:]
A thread signals with \texttt{notify\_all()} before the other enters the wait; the notification is missed despite a guarded \texttt{while} loop.
 See Section~\ref{sec:prob14}

\item[Sleeping Guard:]
A thread goes to sleep on a condition variable without checking the actual shared state, causing missed wakeups and deadlock. See Section~\ref{sec:prob15}

\end{description}

%%%%%%%%%%%%%%%%%%%%%%%%%%%%%%%%%%%%%%%%

\section{Interleaving Multithreaded Code}
\label{sec:simulate}

In this section, we describe our method for simulating multithreaded programs in a controlled and repeatable manner using Python generators. To enable systematic exploration and direct comparison across a variety of concurrency scenarios, we adopt a uniform representation strategy that brings clarity, modularity, and flexibility to our simulation framework.

Each problem is encoded as a collection of Python generator functions. Each generator models a single thread that operates on the System Under Test (SUT) and uses \lstinline{yield} statements to explicitly mark points where execution may pause and control may be transferred to another thread.  Modeling representation allows us to canonize a wide range of concurrency scenarios into a common format, facilitating repeatable experiments and meaningful comparisons under different timing conditions. Our framework further incorporates parameter-dependent delays, which can include both structured variation (e.g., based on thread-specific parameters or environment emulation) and random noise. This enables modeling of both deterministic scheduling and nondeterministic, real-world variability.

Together, these design choices provide a robust and extensible foundation for simulating complex concurrency behaviors and analyzing how timing-related parameters influence system correctness.
The types of problems we address typically involve multiple threads, shared variables, and bugs that are triggered only under specific interleavings, often governed by subtle timing conditions. To simulate such behavior, we employ the \lstinline{simulate()} function shown in \autoref{fig:simulate}, which orchestrates the execution of multiple threads according to a parameter-driven timing model.

\begin{lstlisting}[
caption={Core simulation loop controlling the execution of multiple threads. Each thread yields a delay, and the scheduler selects the next thread to execute based on wake-up times.},
label={fig:simulate},
float=t
]
def simulate(_threads, init=lambda:None, init_arg=None, expected_invariant=None):
    init(init_arg)                         # Initialize global variables
    gen = [t() for t in _threads]          # Create generators (threads)
    wake_times = [0] * len(_threads)       # Initial wake times
    while any(t < END for t in wake_times):
        nxt = np.argmin(wake_times)        # Select next thread to wake
        wake_times[nxt] += next(gen[nxt])  # Advance its wake time
        if expected_invariant is not None:
            assert expected_invariant()    # Check system invariant
\end{lstlisting}

The \lstinline{simulate()} function manages a set of thread generators. Each thread yields a value representing how long it wishes to "sleep" (i.e., delay its next execution step), and the simulation engine schedules the threads based on their wake-up times. The thread with the shortest delay is resumed first, simulating a time-based interleaving of execution steps. Importantly, the simulation does not involve real-time waiting or system-level delays. Instead, it operates in virtual time, advancing the logical clock and reordering thread execution based on the declared delays, thereby allowing efficient exploration of possible interleavings without wasting actual runtime.

Each thread is implemented as a generator function that performs a sequence of atomic operations, with \lstinline{yield} statements marking the boundaries between them. These yield points indicate simulated delays during which other threads may execute. An illustrative example is provided in \autoref{fig:thread}.

\begin{lstlisting}[
caption={A thread modeled as a generator. Yields represent delays between atomic steps. The delays depend on a system-wide coefficient $C$ and problem-specific parameters $D_i$, with optional noise added.},
label={fig:thread},
float=*
]
def simulated_thread():
    global x                            # Shared variable
    for i in range(10):
        yield C * D1 + distortion()     # Simulated delay
        x = 3                           # Atomic operation
    yield C * D2 + distortion()         # Simulated delay
    if x != 3:
        yield C * D3 + distortion()     # Additional delay before assert
        assert x != 3                   # Bug condition
    yield END
\end{lstlisting}

This example models a typical concurrency issue: the thread sets a shared variable \lstinline{x} to a fixed value, but due to interleaved execution, another thread might overwrite it before the current one verifies its value. The timing between steps is simulated by yielding expressions that define how long each thread "sleeps" before proceeding. Each delay expression consists of three components. 
The first is a global coefficient $C$, which reflects the overall processing speed or workload of the simulated system. 
The second is a parameter $D_i$, representing the nominal delay associated with a specific operation. 
The third component is a call to \lstinline{distortion()}, which introduces random variation to simulate environmental unpredictability such as jitter or fluctuating system load.

This parametrization enables the simulation to model a wide range of execution environments and conditions. By adjusting the coefficient $C$, we can emulate machines with varying processing speeds or scheduling overhead. Changing the $D_i$ values allows us to control the logical duration of specific computation segments. The addition of noise via \lstinline{distortion()} allows us to explore nondeterministic interleavings, helping to uncover rare or timing-sensitive bugs that would otherwise be difficult to reproduce.

\textbf{Simulating Rare Failures:} Many concurrency bugs, especially those related to race conditions and ordering violations, are notoriously difficult to reproduce in real systems because they manifest only under rare timing conditions. Our simulation framework addresses this challenge by treating delay parameters as inputs. Specifically, each test-case accepts a tuple of values representing delays (e.g., $D_1$, $D_2$, $D_3$), and runs the simulation multiple times using different random seeds for \lstinline{distortion()}.
Each simulation run returns a result indicating whether a failure (e.g., assertion violation) occurred. By aggregating the outcomes across many runs, we can estimate the probability that a particular configuration of delays leads to a bug. This approach is especially useful for identifying critical thresholds or delay combinations that increase the failure likelihood.

\textbf{Invariant Checking:} Optionally, a predicate \lstinline{expected_invariant} can be passed to the \lstinline{simulate()} function. This predicate is evaluated after each execution step to ensure that the system remains in a valid state. Violations of this invariant are treated as test failures and help pinpoint scenarios of the manifestation of concurrency bugs.

\subsection{Evaluation Protocol}
To enable a fair, consistent, and statistically robust evaluation of the bug-amplification methods under study, we define a controlled experimental framework that governs how test-cases are generated, evaluated, and compared. This framework incorporates a fixed execution budget, multiple randomized trials, and an analysis of the top-performing test-cases across different metrics. Together, these components ensure that our assessment is not only reproducible but also reflective of real-world usage scenarios such as iterative debugging and automated fault localization pipelines.

\textbf{Budget Consumption.} Each bug-amplification method is constrained to a fixed execution budget of~$B$ runs of the SUT. The budget is progressively consumed in increasing blocks of test-case numbers, \(n \in \{100, 300, \dots, 3900\}\), allowing each method to iteratively improve its selection while avoiding early resource exhaustion. At each checkpoint, the accumulated executions are analyzed to update the observed probability of bug exposure. This staged consumption strategy supports convergence analysis and ensures that all methods operate under identical cost constraints while striving to maximize effectiveness. The specific mechanisms by which each method adheres to this budget constraint are detailed in their respective descriptions, and an overview of the budget split across iterations is provided in~\autoref{tab:budget-split}.

\begin{table}[t]
\centering
\caption{Budget allocation per methods. \textbf{brute-force ($BF$)} spends the first \(B/k\) runs, where \(k\) is minimum repeat size, on \emph{estimation} of a random candidate and immediately reports that score; there is no exploitation phase.  \textbf{Simulated Annealing (\textit{SA})} divides the budget into steps \(s\) and neighborhood size \(k\) , enabling explicit control of SUT invocations. The \textbf{Genetic Algorithm (\textit{GA})} uses a population of $k=50$ and evolves for \(B/k\) generations. The \textbf{Ensemble classifier (\textit{Ens})} devotes the entire budget to model‑guided search: at every step, it samples 100 random inputs (exploration) and 100 model‑ranked inputs (exploitation), retrains, and repeats.  }
\label{tab:budget-split}
\begin{tabular}{cccc}
\toprule
\rowcolor{gray!25}
Method & Exploration & Exploitation & Notes\\ \midrule
$BF$ & $B/k$ random & - & k is the minimum repetition required\\
\textit{SA} & $k$ per step & $B/k$ steps & \\
\textit{GA} & pop. $k$ per generation & $B/k$ generations & \\
\textit{Ens} & add 100 random/iter & add 100 ranked/iter & Full budget trains model each iter\\
\bottomrule
\end{tabular}
\end{table}

Nonetheless, while strict budget adherence is maintained throughout each method's execution, the final evaluation phase in this study deliberately exceeds these constraints. This extended phase is not part of the methods themselves, but is introduced solely for the purpose of research evaluation. Specifically, to rigorously assess the quality of the selected test-cases, typically those with the highest observed failure likelihood, we subject each case to massive re-execution with the SUT, far beyond the original budget. This allows us to derive a precise and statistically robust estimate of its true failure-inducing potential.

\textbf{Repeated trials.}  To obtain statistically meaningful estimates, each $\langle method, problem \rangle$ pair was executed \(50\) independent times.  Each run exploited the full budget schedule above, producing a \emph{single} best‑scoring test-case, i.e., the input with the highest observed failure probability.  Aggregating the best scores over 50 runs yields the sample mean and standard deviation that appear in all result plots.

\textbf{Top-$k$ Analysis.} In practice, automated debugging pipelines require three disjoint pools: development (\textit{debugging}), model training (\textit{testing}), and final assessment (\textit{validation}).  Reporting only the single best-case risks overfits, whereas presenting the entire budget is often impractical.  Hence, we also study the $5^{\text{th}}$ and $10^{\text{th}}$ best inputs, providing a small yet diverse set that effectively supports such a pipeline.

%%%%%%%%%%%%%%%%%%%%%%%%%%%%%%%%%%%%%%%%%%%%%%%%%%%%%%%%%%%%%%%%%

\section{Bug-Amplification Methods}
\label{sec:tested-approaches}
This section presents the various search techniques explored in our study to amplify the probability of detecting concurrency bugs. The generation of effective test-cases presents both a statistical and an algorithmic challenge. Our goal is to investigate whether advanced heuristics can outperform naive or exhaustive methods in this context. Each subsection below introduces a distinct test generation paradigm, ranging from brute-force enumeration to learning-based classification, and describes its design, rationale, and implementation as applied to our concurrency benchmark. 

\subsection{Baseline: Random Search}
\label{sec:subBF}
Random search serves as the baseline method in this study, providing a critical comparison point for evaluating the effectiveness of more sophisticated search techniques. This method operates without incorporating any domain-specific heuristics or optimization strategies, offering conceptual simplicity and ease of implementation. Its role is to help determine whether complex methods are truly necessary, or if random exploration is sufficient for discovering high-probability failure-inducing test-cases.

The process begins by randomly generating $(B/k)$ candidate test-cases. Each candidate is evaluated by executing it multiple $(k)$ times against the system under test, in order to estimate its likelihood of triggering a bug. Every execution yields a binary outcome -- failure (bug-triggered) or success. A scenario's estimated score is calculated as the frequency of failures across its executions, with the number of repetitions serving as a configurable parameter that trades evaluation accuracy for computational cost.

We used a fixed sampling parameter $k = 30$ for each test-case. This value was chosen based on the statistical justification provided by the Law of Large Numbers (LLN) and the Central Limit Theorem (CLT), which suggest that 30 independent samples are generally sufficient to obtain a stable estimate of the mean and variance. This ensures that the bug exposure probability computed from the $k$ test-cases is both statistically meaningful and computationally efficient.

Like all tested methods, random search operates within a fixed execution budget $B$ as described earlier. 
Once the budget is consumed, candidates are ranked by their estimated failure probability, and the top-ranked scenarios are selected as the method's output. Throughout the remainder of the paper, we refer to this approach as the Brute-Force ($BF$) method.

In the Results section, we examine scenarios where advanced search methods offer clear benefits and compare them with cases where the $BF$ method performs adequately.

%------------------------------------------------------
\subsection{Simulated Annealing\label{subsec:sa-heuristic}}

Finding a concurrency bug in a continuous search space can be viewed as
climbing an unknown, locally smooth \emph{probability landscape}
$p(x)$ whose height at a point $x\!\in\!\mathbb{R}^{n}$ represents the
likelihood that the corresponding test input triggers the fault.
Our Simulated-Annealing (\textit{SA}) variant explores this landscape by iteratively sampling a small neighborhood of the current point and then moving in the direction where failures are more concentrated.

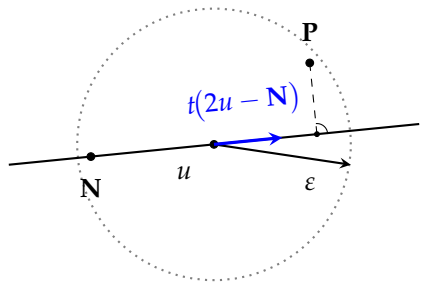
\begin{figure}[H]
  \centering
  \begin{tikzpicture}[>=stealth,scale=.9]
    % --- frame ---------------------------------------------------
    \node[thick,rounded corners,inner sep=8pt]{
      \begin{tikzpicture}[scale=.9]
        % --- enclosing ball -------------------------------------
        \draw[dotted, thick, gray] (0,0) circle (2);   % radius ε (not to scale)

        % --- radius arrow and label ε ---------------------------
        \draw[->, thick] (0,0) -- (2,-0.3) node[midway, below right] {$\varepsilon$};

        % --- main search line through N and u -------------------
        \coordinate (AA) at (-3,-0.3);
        \coordinate (BB) at ( 3, 0.3);          % slope = 0.1
        \draw[thick] (AA) -- (BB);

        % --- points N and u -------------------------------------
        \coordinate (N) at (-1.8,-0.18);
        \coordinate (u) at ( 0,  0   );
        \filldraw (N) circle (1.6pt) node[below] {$\mathbf N$};
        \filldraw (u) circle (1.6pt) node[below left] {$u$};

        % --- positive center P ----------------------------------
        \coordinate (P) at ( 1.4, 1.2 );
'   \filldraw (P) circle (1.6pt) node[above] {$\mathbf P$};

        % --- orthogonal projection of P onto search line --------
        \coordinate (Q) at ( 1.505, 0.1505 );
        \filldraw (Q) circle (1.0pt);
        \draw[dashed] (P) -- (Q);

        % --- right-angle tick -----------------------------------
        \coordinate (R) at ($(Q)+(1,0.1)$);
        \pic [draw,angle radius=4pt] {right angle = P--Q--R};

        % --- movement arrow -------------------------------------
        \draw[->,very thick,blue]
              (u) -- (1,0.1)
              node[midway,above,sloped] {$t\!\bigl(2u-\mathbf N\bigr)$};
      \end{tikzpicture}
    };
  \end{tikzpicture}
  \caption{Geometric intuition in 2D of the update step.
           The $k$ candidates are sampled inside the dotted ball
           $B(u,\varepsilon)$ centered at $u$.
           We move from $u$ to
           $u_{\text{next}} = t\bigl(2u-\mathbf N\bigr)$ (blue arrow) and,
           if positives exist, bias the step toward the positive center
           $\mathbf P$. The dashed segment from $\mathbf P$ is perpendicular
           to the search line. The radius of the sampling ball is marked by
           $\varepsilon$.}
  \label{fig:sa-heuristic}
\end{figure}

\textbf{Why this variant?}
We developed this \textit{SA} variant for three reasons.
(i)~\textbf{Budget control:}  By fixing $k$ candidates per step ($k=30$ as described in \ref{sec:subBF}) and $s$ ($s=B/k$) optimization steps, we guarantee an exact run budget $B$. Most generic \textit{SA} frameworks expose only the iteration count and can silently overshoot the allowed SUT executions.
(ii)~\textbf{Noise awareness.}  Each fitness evaluation is stochastic, so the algorithm must cope with noisy measurements, a feature rarely found in off-the-shelf \textit{SA} libraries.
(iii)~\textbf{Geometric clarity.}  The center-of-mass update rule (see \autoref{fig:sa-heuristic}) offers an intuitive, easily inspectable implementation that has proven effective in practice.

\textbf{Neighbourhood sampling.} Let $u\!\in\!\mathbb{R}^{n}$ be the current input vector.
We draw $k$ random candidates $\{x_{1},\dots,x_{k}\}$ from the ball $B(u,\varepsilon)=\{x\mid\lVert x-u\rVert\le\varepsilon\}$.
After executing each candidate, we label it \emph{positive} ($\mathrm P$) if it triggers the bug $(p(x_{i})=1)$ and \emph{negative} ($\mathrm N$) otherwise $(p(x_{i})=0)$.

\textbf{Center-of-mass estimate.} We summarize the neighborhood by the averages
\[
  \mathbf N=\frac{1}{|\mathrm N|}\sum_{x_{i}\in\mathrm N}x_{i},
  \qquad
  \mathbf P=\frac{1}{|\mathrm P|}\sum_{x_{i}\in\mathrm P}x_{i},
\]
which act as coarse estimates of where the bug is less
($\mathbf N$) or more ($\mathbf P$) likely to occur.

\textbf{Update rule.}
Intuitively, we wish to move away from the negatives and, if positives
exist, steer toward their center of mass.
We therefore (i) take a step from $u$ opposite to $\mathbf N$ and
(ii)~if positives are present, average that tentative step with 
$\mathbf P$'s position.
The construction is illustrated in  \autoref{fig:sa-heuristic}, and a
Python sketch appears in \autoref{fig:AS_app}.

\textbf{Edge cases:}
If no positive points or no negative points are found, we choose $u_{\text{next}}$ as a random point within $B(u,\varepsilon)$.
As $\varepsilon$ gradually decreases (the usual annealing schedule), the
search converges on increasingly precise regions of high failure
probability while still escaping unpromising basins.

\begin{lstlisting}[
caption={An implementation of Neighbourhoods Sampling.},
label={fig:AS_app},
float=*
]
def next_point(u, epsilon=0.1, k=30, bounds=[(0,1)]*20):
    # Step 1: Randomly choose k points in the ball B(u, epsilon)
    S = [generate_within_bounds(u, epsilon, bounds) for _ in range(k)]

    # Step 2: Execute each x_i and determine whether the bug was found
    id_x = [run_test(np.array(x_i)) for x_i in S]

    # Step 3: Create two averages N and P
    N = np.mean([x_i for x_i, id in zip(S, id_x) if id == 0], axis=0)
    P = np.mean([x_i for x_i, id in zip(S, id_x) if id == 1], axis=0)

    if P.empty() or N.empty():
        u_next =  S[0]     # arbitrary point
    else:
        # Step 4: Obtain a new point w' and take the average of P and w' as the next point in the search
        w_prime = 2*u - N
        u_next = (P + w_prime) / 2

    return u_next
\end{lstlisting}    

\subsection{Genetic Algorithm-Based Search}

To explore failure-inducing test-cases, we employed a Genetic Algorithm (\textit{GA}) using the EC‑KitY evolutionary computation framework~\cite{sipper2023eckity}. The goal of the algorithm is to evolve test inputs that are likely to trigger failures in a concurrent system, guided by a fitness function that reflects the probability of failure.

We configured the \textit{GA} with a population size of $k=50$ individuals per generation. The total number of generations is determined by dividing the available test-cases budget by the population size, ensuring that each individual is evaluated once per generation. Fitness is computed using a user-defined \texttt{BugHuntingEvaluator}, which estimates the likelihood of a bug manifesting during execution. Since this is a maximization task, higher fitness indicates more failure-prone test-cases.
Each individual is represented as a real-valued vector constrained within predefined bounds (depending on each problem). 

%--------------------------

We tuned the genetic algorithm's hyperparameters to balance convergence speed and search diversity while staying within our evaluation budget. To that end, we selected a population size of $k=50$, following classical guidelines by Goldberg~\cite{goldberg1989} and more recent studies~\cite{Karafotias2015} that recommend sizes in the range of 30--100 to ensure sufficient diversity without excessive cost. We used a two-point crossover with a 0.5 probability to promote recombination of substructures, and uniform mutation applied to 10 randomly selected components with a 0.15 probability to inject controlled variation. Tournament selection with a size of four was chosen to provide moderate selective pressure while preserving population diversity. These values were selected based on standard practice and empirical effectiveness in evolutionary search.

%---------------------------

The \textit{GA} employs the following operators:

\begin{itemize}
    \item \textbf{Crossover:} A two-point crossover (\texttt{VectorKPointsCrossover}) with a probability of 0.5 exchanges two genome segments between parent individuals. This promotes the recombination of useful substructures and accelerates convergence.
    
    \item \textbf{Mutation:} Uniform N-point mutation (\texttt{FloatVectorUniformNPointMutation}) is applied to 10 randomly selected vector components with a probability of 0.15. This introduces variation and helps the population explore new regions in the search space.
    
    \item \textbf{Selection:} We use tournament selection with a size of four, where the fittest individual among four randomly sampled candidates is chosen as a parent. This balances selective pressure and population diversity.
\end{itemize}

We applied elitism by retaining the single best individual in each generation, and terminated the run if no improvement was observed in the best fitness over 100 consecutive generations.

The EC-KitY is a modular and extensible evolutionary computation toolkit for Python, designed to support a wide range of evolutionary techniques including genetic algorithms, genetic programming, coevolution, and multi-objective optimization. It also provides seamless integration with machine learning pipelines, particularly via \texttt{scikit-learn}.

 As part of our investigation into effective methods for test-case generation, we explored using  Genetic Programming, based on the hypothesis that dependencies exist among the input parameters of failure-inducing scenarios. Specifically, we considered the possibility of defining a domain-specific language capable of capturing structural patterns and relations between parameters that frequently lead to failures. This hypothesis was inspired by prior work suggesting that, in most cases, only a small subset of input parameters is responsible for triggering bugs~\cite{Elyasaf2023}.

However, despite initial efforts, the genetic programming process failed to converge toward meaningful patterns, and the approach was ultimately abandoned. As a result, we redirected our efforts toward a more conventional search method, utilizing a generic Genetic Algorithm (\textit{GA}) instead.

%xxxxxxxxxxxxxxxxxxxxxxxxxxxxxxxxxxxxxxxxxxxxxxxxxxxxxxxxxxxxxxxx

\subsection{Classification-Based Method: Ensemble Stacking Classifier}

In this study, we investigated several supervised learning techniques to enhance the identification of failure, inducing test-cases. After evaluating multiple classifiers, including Random Forests and Multilayer Perceptrons, we observed no significant differences in performance between them. In addition, we experimented with several regression-based models, but they failed to provide reliable prioritization of failure-prone test-cases. As a result, we adopted the Ensemble Stacking Classifier as our primary model, leveraging its ability to combine the strengths of various base learners into a unified predictive framework.

%------------------------------------------------------
\paragraph{Stacking Architecture}
\textit{Layer 1} comprises four diverse classifiers-Logistic Regression, Decision Tree, Random Forest, and an MLP, each trained independently to return the probability that a test-case triggers a failure.  
\textit{Layer~2} is a Logistic Regression meta-learner that ingests both the base-model probabilities and the raw input features (via \texttt{passthrough = True}).  
To curb overfitting, the meta-learner is trained with 5-fold cross-validation, using out-of-fold predictions from the base models. \autoref{fig:Ensamble} presents the Ensemble Stacking Classification implementation, using the most common classifiers together.

%------------------------------------------------------
\textbf{Pre-processing:}
To address the inherent class imbalance in our data, we apply the \textit{Synthetic Minority Over-sampling Technique} (SMOTE) before training. This ensures a balanced representation of failure and non-failure cases, which improves generalization and stabilizes model training. After training, the ensemble classifier assigns a failure probability to each unseen test-case. These predictions are used to rank test-cases, enabling prioritized execution under a limited testing budget, with failure-prone inputs examined first.

\textbf{Data Preparation Pipeline:}
To ensure effective training of the ensemble model, we employed a structured data preparation pipeline comprising three integrated phases. The process began with an initialization step, where we seeded the training dataset using test-cases that had previously triggered system failures during the early stages of bug discovery. This provided a foundational set of informative examples for the model to learn from.

In the next phase, we extended the dataset through a budget-guided expansion strategy. This included both exploitation and exploration mechanisms: the model was used to identify new test-cases with high predicted failure probabilities (exploitation), while additional test-cases were also sampled randomly (exploration) to ensure input diversity and guard against model bias.

Finally, in the evaluation phase, the trained model was applied to a large pool of previously unseen test inputs. Based on the predicted failure likelihood, we selected the top-ranked cases for exhaustive system execution. This allowed us to assess the actual failure rates of prioritized inputs, independently of the training budget, thereby providing a robust estimate of the model's predictive utility.
\begin{lstlisting}[
caption={The Ensemble Stacking Classifier. After experimentation, we found that stacking the four most common classifiers and combining their predictions using logistic regression gives the best results. We configured \texttt{passthrough=True} to allow raw features to reach the meta-learner and \texttt{cv=5} for robust out-of-fold training. We also adjusted the number of iterations to cope with model complexity and used \texttt{class\_weight='balanced'} due to skewed data, as bugs are rarely triggered. A two-layer neural network with adaptive learning rate further enhances abstraction and generalization.},
label={fig:Ensamble},
float=*
]
base_learners = [
    ('lr', LogisticRegression(max_iter=1000, class_weight='balanced')),
    ('dt', DecisionTreeClassifier(class_weight='balanced')),
    ('rf', RandomForestClassifier(n_estimators=100, class_weight='balanced')),
    ('mlp', MLPClassifier(hidden_layer_sizes=(50, 20), 
            learning_rate='adaptive', max_iter=500, early_stopping=True))
]

meta_learner = LogisticRegression(class_weight='balanced',max_iter=1000)

stacked_model = StackingClassifier(
    estimators=base_learners,
    final_estimator=meta_learner,
    cv=5,
    passthrough=True)
\end{lstlisting}

\textbf{Observations and Rationale:}
The Ensemble Stacking Classifier consistently demonstrated reliable and accurate predictions across experiments, providing a balanced trade-off between generalization, robustness, and computational feasibility. Its ability to incorporate multiple perspectives from heterogeneous learners contributed to a more stable and accurate prioritization of test-cases. This made it a natural choice as the core classification model in our failure detection framework.
%xxxxxxxxxxxxxxxxxxxxxxxxxxxxxxxxxxxxxxxxxxxxxxxxxxxxxxxxxxxxxxxxxxxxxxxxxxx

%======================================================
\section{Results}\label{sec:results}
%======================================================

This section presents the empirical results obtained from evaluating our test generation framework on a suite of 17 benchmark concurrency problems, each containing a known, seeded bug. To assess effectiveness, we applied four black-box test generation methods: \textit{Brute-Force ($BF$)}, \textit{Ensemble Classifier (\textit{Ens})}, \textit{Genetic Algorithm (\textit{GA})}, and \textit{Simulated Annealing (\textit{SA})}. Each method was executed in 50 independent trials per problem to mitigate the influence of stochastic variability and enable robust statistical analysis.
During each execution, the method generated a unique test suite, and we recorded whether any test-case within it successfully triggered the target bug. For every method–problem pair, we computed the empirical probability of failure discovery, alongside standard deviation and 95\% confidence intervals.

The structure of this section follows the types of visualizations used to interpret the results: overall bug-detection rates per problem, convergence behavior as a function of test budget, top-ranked test-case performance comparisons, and statistical significance analyses between methods. Each type of graph is introduced with an explanation of what it conveys, how the data is structured, and what insights emerge from the results.

\subsection{Overall Success Rates per Problem}

This subsection presents the average probability of successfully triggering each bug using the four tested methods. Each bar represents the mean probability computed across 50 independent runs for a fixed number of test-cases. 

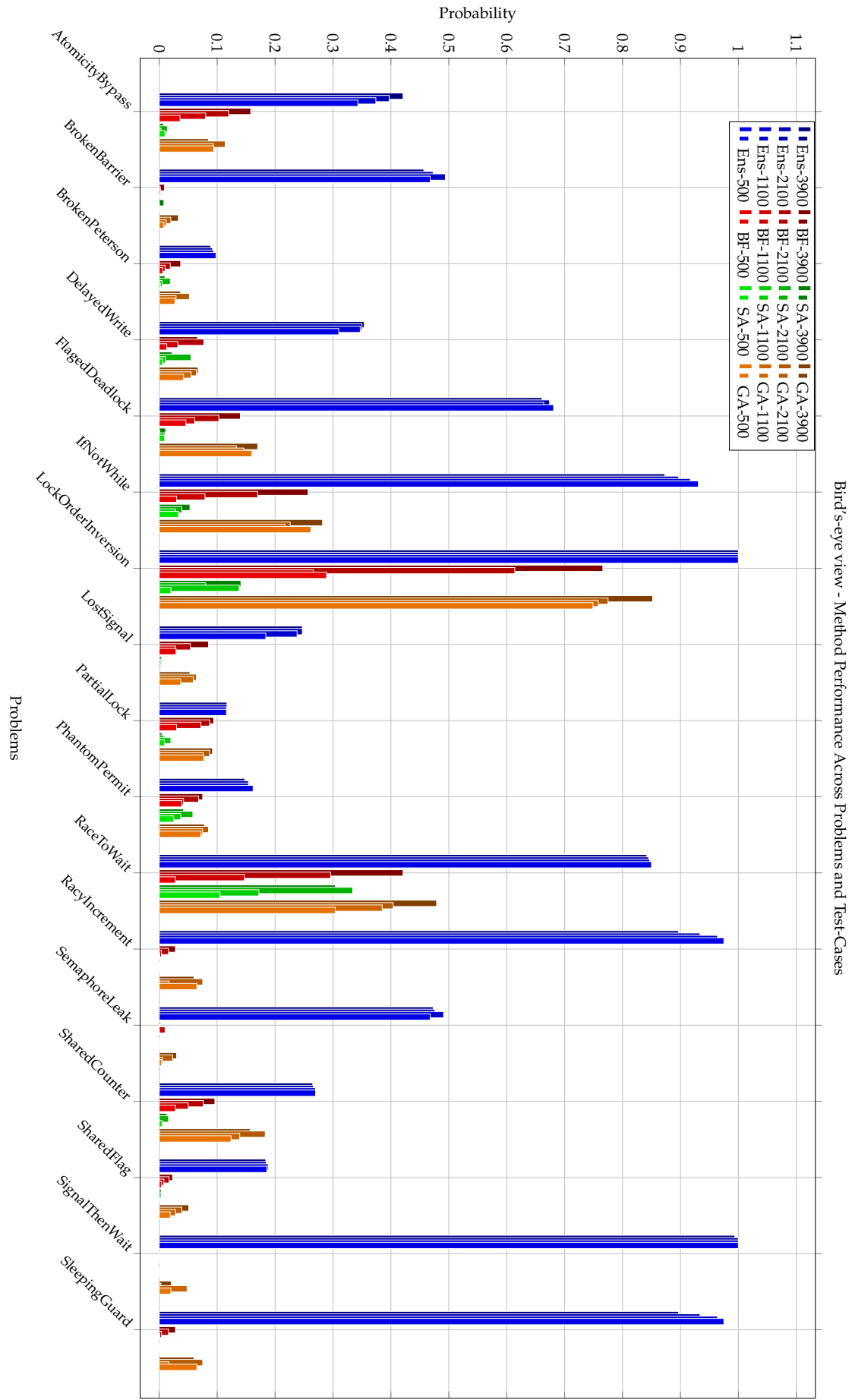
\begin{figure}
\centering
\rotatebox{270}{
\scalebox{0.78}{
\begin{tikzpicture}
\begin{axis}[
    ybar,
    bar width=4pt,
    width=30cm,
    height=15cm,
    symbolic x coords={AtomicityBypass-Ens, AtomicityBypass-BF, AtomicityBypass-SA, AtomicityBypass-GA, spacer1, BrokenBarrier-Ens, BrokenBarrier-BF, BrokenBarrier-SA, BrokenBarrier-GA, spacer2, BrokenPeterson-Ens, BrokenPeterson-BF, BrokenPeterson-SA, BrokenPeterson-GA, spacer3, DelayedWrite-Ens, DelayedWrite-BF, DelayedWrite-SA, DelayedWrite-GA, spacer4, FlagedDeadlock-Ens, FlagedDeadlock-BF, FlagedDeadlock-SA, FlagedDeadlock-GA, spacer5, IfNotWhile-Ens, IfNotWhile-BF, IfNotWhile-SA, IfNotWhile-GA, spacer6, LockOrderInversion-Ens, LockOrderInversion-BF, LockOrderInversion-SA, LockOrderInversion-GA, spacer7, LostSignal-Ens, LostSignal-BF, LostSignal-SA, LostSignal-GA, spacer8, PartialLock-Ens, PartialLock-BF, PartialLock-SA, PartialLock-GA, spacer9, PhantomPermit-Ens, PhantomPermit-BF, PhantomPermit-SA, PhantomPermit-GA, spacer10, RaceToWait-Ens, RaceToWait-BF, RaceToWait-SA, RaceToWait-GA, spacer11, RacyIncrement-Ens, RacyIncrement-BF, RacyIncrement-SA, RacyIncrement-GA, spacer12, SemaphoreLeak-Ens, SemaphoreLeak-BF, SemaphoreLeak-SA, SemaphoreLeak-GA, spacer13, SharedCounter-Ens, SharedCounter-BF, SharedCounter-SA, SharedCounter-GA, spacer14, SharedFlag-Ens, SharedFlag-BF, SharedFlag-SA, SharedFlag-GA, spacer15, SignalThenWait-Ens, SignalThenWait-BF, SignalThenWait-SA, SignalThenWait-GA, spacer16, SleepingGuard-Ens, SleepingGuard-BF, SleepingGuard-SA, SleepingGuard-GA},
    xtick={AtomicityBypass-BF, BrokenBarrier-BF, BrokenPeterson-BF, DelayedWrite-BF, FlagedDeadlock-BF, IfNotWhile-BF, LockOrderInversion-BF, LostSignal-BF, PartialLock-BF, PhantomPermit-BF, RaceToWait-BF, RacyIncrement-BF, SemaphoreLeak-BF, SharedCounter-BF, SharedFlag-BF, SignalThenWait-BF, SleepingGuard-BF},
    xticklabels={AtomicityBypass, BrokenBarrier, BrokenPeterson, DelayedWrite, FlagedDeadlock, IfNotWhile, LockOrderInversion, LostSignal, PartialLock, PhantomPermit, RaceToWait, RacyIncrement, SemaphoreLeak, SharedCounter, SharedFlag, SignalThenWait, SleepingGuard},
    x tick label style={rotate=45, anchor=east},
    xlabel={Problems},
    ylabel={Probability},
    title={Bird's-eye view - Method Performance Across Problems and Test-Cases},
    legend style={at={(0.17,0.999))},anchor=north,legend columns=4},
    ymin=0,
    ymax=1.1,
    grid=major,
    enlargelimits=0.03,
    scale only axis=true, 
    xticklabel style={font=\small}, % or \footnotesize, \scriptsize, \tiny, \large, etc.
]

\addplot+[
    bar shift=0pt,
    draw=white,
    fill=black!50!blue
] coordinates {
    (AtomicityBypass-Ens, 0.421) (BrokenBarrier-Ens, 0.457) (BrokenPeterson-Ens, 0.089) (DelayedWrite-Ens, 0.354) (FlagedDeadlock-Ens, 0.661) (IfNotWhile-Ens, 0.873) (LockOrderInversion-Ens, 1.000) (LostSignal-Ens, 0.247) (PartialLock-Ens, 0.117) (PhantomPermit-Ens, 0.148) (RaceToWait-Ens, 0.842) (RacyIncrement-Ens, 0.897) (SemaphoreLeak-Ens, 0.473) (SharedCounter-Ens, 0.265) (SharedFlag-Ens, 0.184) (SignalThenWait-Ens, 0.994) (SleepingGuard-Ens, 0.897)
};

\addplot+[
    bar shift=0pt,
    draw=white,
    fill=black!50!red
] coordinates {
    (AtomicityBypass-BF, 0.158) (BrokenBarrier-BF, 0.009) (BrokenPeterson-BF, 0.037) (DelayedWrite-BF, 0.066) (FlagedDeadlock-BF, 0.140) (IfNotWhile-BF, 0.257) (LockOrderInversion-BF, 0.766) (LostSignal-BF, 0.085) (PartialLock-BF, 0.094) (PhantomPermit-BF, 0.075) (RaceToWait-BF, 0.421) (RacyIncrement-BF, 0.028) (SemaphoreLeak-BF, 0.002) (SharedCounter-BF, 0.096) (SharedFlag-BF, 0.023) (SignalThenWait-BF, 0.000) (SleepingGuard-BF, 0.028)
};

\addplot+[
    bar shift=0pt,
    draw=white,
    fill=black!50!green
] coordinates {
    (AtomicityBypass-SA, 0.008) (BrokenBarrier-SA, 0.008) (BrokenPeterson-SA, 0.010) (DelayedWrite-SA, 0.022) (FlagedDeadlock-SA, 0.011) (IfNotWhile-SA, 0.053) (LockOrderInversion-SA, 0.141) (LostSignal-SA, 0.004) (PartialLock-SA, 0.005) (PhantomPermit-SA, 0.042) (RaceToWait-SA, 0.304) (RacyIncrement-SA, 0.000) (SemaphoreLeak-SA, 0.000) (SharedCounter-SA, 0.013) (SharedFlag-SA, 0.004) (SignalThenWait-SA, 0.000) (SleepingGuard-SA, 0.000)
};

\addplot+[
    bar shift=0pt,
    draw=white,
    fill=black!50!orange
] coordinates {
    (AtomicityBypass-GA, 0.085) (BrokenBarrier-GA, 0.033) (BrokenPeterson-GA, 0.037) (DelayedWrite-GA, 0.067) (FlagedDeadlock-GA, 0.170) (IfNotWhile-GA, 0.282) (LockOrderInversion-GA, 0.852) (LostSignal-GA, 0.053) (PartialLock-GA, 0.092) (PhantomPermit-GA, 0.078) (RaceToWait-GA, 0.479) (RacyIncrement-GA, 0.060) (SemaphoreLeak-GA, 0.030) (SharedCounter-GA, 0.157) (SharedFlag-GA, 0.051) (SignalThenWait-GA, 0.021) (SleepingGuard-GA, 0.060)
};

\addplot+[
    bar shift=1.5pt,
    draw=white,
    fill=black!30!blue
] coordinates {
    (AtomicityBypass-Ens, 0.397) (BrokenBarrier-Ens, 0.473) (BrokenPeterson-Ens, 0.092) (DelayedWrite-Ens, 0.350) (FlagedDeadlock-Ens, 0.674) (IfNotWhile-Ens, 0.897) (LockOrderInversion-Ens, 1.000) (LostSignal-Ens, 0.247) (PartialLock-Ens, 0.117) (PhantomPermit-Ens, 0.154) (RaceToWait-Ens, 0.845) (RacyIncrement-Ens, 0.934) (SemaphoreLeak-Ens, 0.476) (SharedCounter-Ens, 0.266) (SharedFlag-Ens, 0.185) (SignalThenWait-Ens, 1.000) (SleepingGuard-Ens, 0.934)
};

\addplot+[
    bar shift=1.5pt,
    draw=white,
    fill=black!30!red
] coordinates {
    (AtomicityBypass-BF, 0.120) (BrokenBarrier-BF, 0.002) (BrokenPeterson-BF, 0.019) (DelayedWrite-BF, 0.077) (FlagedDeadlock-BF, 0.103) (IfNotWhile-BF, 0.170) (LockOrderInversion-BF, 0.614) (LostSignal-BF, 0.054) (PartialLock-BF, 0.087) (PhantomPermit-BF, 0.068) (RaceToWait-BF, 0.296) (RacyIncrement-BF, 0.016) (SemaphoreLeak-BF, 0.001) (SharedCounter-BF, 0.076) (SharedFlag-BF, 0.017) (SignalThenWait-BF, 0.000) (SleepingGuard-BF, 0.016)
};

\addplot+[
    bar shift=1.5pt,
    draw=white,
    fill=black!30!green
] coordinates {
    (AtomicityBypass-SA, 0.014) (BrokenBarrier-SA, 0.000) (BrokenPeterson-SA, 0.019) (DelayedWrite-SA, 0.055) (FlagedDeadlock-SA, 0.001) (IfNotWhile-SA, 0.039) (LockOrderInversion-SA, 0.080) (LostSignal-SA, 0.002) (PartialLock-SA, 0.008) (PhantomPermit-SA, 0.058) (RaceToWait-SA, 0.334) (RacyIncrement-SA, 0.001) (SemaphoreLeak-SA, 0.000) (SharedCounter-SA, 0.016) (SharedFlag-SA, 0.003) (SignalThenWait-SA, 0.000) (SleepingGuard-SA, 0.001)
};

\addplot+[
    bar shift=1.5pt,
    draw=white,
    fill=black!30!orange
] coordinates {
    (AtomicityBypass-GA, 0.114) (BrokenBarrier-GA, 0.020) (BrokenPeterson-GA, 0.052) (DelayedWrite-GA, 0.064) (FlagedDeadlock-GA, 0.133) (IfNotWhile-GA, 0.226) (LockOrderInversion-GA, 0.775) (LostSignal-GA, 0.064) (PartialLock-GA, 0.087) (PhantomPermit-GA, 0.085) (RaceToWait-GA, 0.404) (RacyIncrement-GA, 0.075) (SemaphoreLeak-GA, 0.023) (SharedCounter-GA, 0.183) (SharedFlag-GA, 0.039) (SignalThenWait-GA, 0.003) (SleepingGuard-GA, 0.075)
};

\addplot+[
    bar shift=3pt,
    draw=white,
    fill=black!20!blue
] coordinates {
    (AtomicityBypass-Ens, 0.374) (BrokenBarrier-Ens, 0.494) (BrokenPeterson-Ens, 0.094) (DelayedWrite-Ens, 0.347) (FlagedDeadlock-Ens, 0.664) (IfNotWhile-Ens, 0.917) (LockOrderInversion-Ens, 1.000) (LostSignal-Ens, 0.238) (PartialLock-Ens, 0.116) (PhantomPermit-Ens, 0.154) (RaceToWait-Ens, 0.847) (RacyIncrement-Ens, 0.964) (SemaphoreLeak-Ens, 0.491) (SharedCounter-Ens, 0.270) (SharedFlag-Ens, 0.188) (SignalThenWait-Ens, 1.000) (SleepingGuard-Ens, 0.964)
};

\addplot+[
    bar shift=3pt,
    draw=white,
    fill=black!20!red
] coordinates {
        (AtomicityBypass-BF, 0.080) (BrokenBarrier-BF, 0.002) (BrokenPeterson-BF, 0.010) (DelayedWrite-BF, 0.032) (FlagedDeadlock-BF, 0.061) (IfNotWhile-BF, 0.079) (LockOrderInversion-BF, 0.266) (LostSignal-BF, 0.028) (PartialLock-BF, 0.072) (PhantomPermit-BF, 0.042) (RaceToWait-BF, 0.147) (RacyIncrement-BF, 0.004) (SemaphoreLeak-BF, 0.010) (SharedCounter-BF, 0.050) (SharedFlag-BF, 0.008) (SignalThenWait-BF, 0.000) (SleepingGuard-BF, 0.004)
};

\addplot+[
    bar shift=3pt,
    draw=white,
    fill=black!20!green
] coordinates {
    (AtomicityBypass-SA, 0.003) (BrokenBarrier-SA, 0.000) (BrokenPeterson-SA, 0.005) (DelayedWrite-SA, 0.011) (FlagedDeadlock-SA, 0.010) (IfNotWhile-SA, 0.028) (LockOrderInversion-SA, 0.138) (LostSignal-SA, 0.003) (PartialLock-SA, 0.020) (PhantomPermit-SA, 0.037) (RaceToWait-SA, 0.172) (RacyIncrement-SA, 0.000) (SemaphoreLeak-SA, 0.000) (SharedCounter-SA, 0.000) (SharedFlag-SA, 0.000) (SignalThenWait-SA, 0.000) (SleepingGuard-SA, 0.000)
};

\addplot+[
    bar shift=3pt,
    draw=white,
    fill=black!20!orange
] coordinates {
    (AtomicityBypass-GA, 0.093) (BrokenBarrier-GA, 0.012) (BrokenPeterson-GA, 0.029) (DelayedWrite-GA, 0.055) (FlagedDeadlock-GA, 0.146) (IfNotWhile-GA, 0.217) (LockOrderInversion-GA, 0.758) (LostSignal-GA, 0.059) (PartialLock-GA, 0.076) (PhantomPermit-GA, 0.075) (RaceToWait-GA, 0.385) (RacyIncrement-GA, 0.017) (SemaphoreLeak-GA, 0.006) (SharedCounter-GA, 0.139) (SharedFlag-GA, 0.028) (SignalThenWait-GA, 0.048) (SleepingGuard-GA, 0.017)
};

\addplot+[
    bar shift=4.5pt,
    draw=white,
    fill=black!10!blue
] coordinates {
    (AtomicityBypass-Ens, 0.343) (BrokenBarrier-Ens, 0.468) (BrokenPeterson-Ens, 0.098) (DelayedWrite-Ens, 0.310) (FlagedDeadlock-Ens, 0.681) (IfNotWhile-Ens, 0.931) (LockOrderInversion-Ens, 1.000) (LostSignal-Ens, 0.184) (PartialLock-Ens, 0.116) (PhantomPermit-Ens, 0.162) (RaceToWait-Ens, 0.850) (RacyIncrement-Ens, 0.975) (SemaphoreLeak-Ens, 0.468) (SharedCounter-Ens, 0.270) (SharedFlag-Ens, 0.186) (SignalThenWait-Ens, 1.000) (SleepingGuard-Ens, 0.975)
};

\addplot+[
    bar shift=4.5pt,
    draw=white,
    fill=black!10!red
] coordinates {
        (AtomicityBypass-BF, 0.036) (BrokenBarrier-BF, 0.000) (BrokenPeterson-BF, 0.006) (DelayedWrite-BF, 0.013) (FlagedDeadlock-BF, 0.046) (IfNotWhile-BF, 0.030) (LockOrderInversion-BF, 0.289) (LostSignal-BF, 0.029) (PartialLock-BF, 0.030) (PhantomPermit-BF, 0.039) (RaceToWait-BF, 0.028) (RacyIncrement-BF, 0.001) (SemaphoreLeak-BF, 0.000) (SharedCounter-BF, 0.028) (SharedFlag-BF, 0.004) (SignalThenWait-BF, 0.000) (SleepingGuard-BF, 0.001)
};

\addplot+[
    bar shift=4.5pt,
    draw=white,
    fill=black!10!green
] coordinates {
    (AtomicityBypass-SA, 0.010) (BrokenBarrier-SA, 0.000) (BrokenPeterson-SA, 0.002) (DelayedWrite-SA, 0.006) (FlagedDeadlock-SA, 0.009) (IfNotWhile-SA, 0.033) (LockOrderInversion-SA, 0.020) (LostSignal-SA, 0.002) (PartialLock-SA, 0.009) (PhantomPermit-SA, 0.025) (RaceToWait-SA, 0.105) (RacyIncrement-SA, 0.000) (SemaphoreLeak-SA, 0.000) (SharedCounter-SA, 0.005) (SharedFlag-SA, 0.000) (SignalThenWait-SA, 0.000) (SleepingGuard-SA, 0.000)
};

\addplot+[
    bar shift=4.5pt,
    draw=white,
    fill=black!10!orange
] coordinates {
    (AtomicityBypass-GA, 0.094) (BrokenBarrier-GA, 0.008) (BrokenPeterson-GA, 0.027) (DelayedWrite-GA, 0.042) (FlagedDeadlock-GA, 0.160) (IfNotWhile-GA, 0.262) (LockOrderInversion-GA, 0.749) (LostSignal-GA, 0.037) (PartialLock-GA, 0.077) (PhantomPermit-GA, 0.072) (RaceToWait-GA, 0.304) (RacyIncrement-GA, 0.065) (SemaphoreLeak-GA, 0.004) (SharedCounter-GA, 0.124) (SharedFlag-GA, 0.019) (SignalThenWait-GA, 0.020) (SleepingGuard-GA, 0.065)
};
\legend{Ens-3900, BF-3900, SA-3900, GA-3900, Ens-2100, BF-2100, SA-2100, GA-2100, Ens-1100, BF-1100, SA-1100, GA-1100, Ens-500, BF-500, SA-500, GA-500,}

\end{axis}
\end{tikzpicture}
    }
}
\caption{Bird's-eye view for all problems, probability of triggering bug after 500, 1100, 2100, and 3900 test-cases. Each bar is one experiment and based on 50 independent runs. The X axis is all 17 problems, and for each problem, 4 methods and 4 (out of 20) test-cases are shown. The y-axis is the maximum probability for the best test-case.}
\label{fig:OVall}
\end{figure}

The graph in \autoref{fig:OVall} allows direct comparison of method effectiveness across the 17 problems. As observed for 500, 1100, 2100, and 3900 test-cases, the \textit{Ens} method consistently outperforms the other methods in most cases, often achieving significantly higher success rates with lower variance. The $BF$ method generally lags behind, especially on more complex bugs.

Across all 17 benchmarks, the ensemble (\textit{Ens}) already achieves a mean success probability of $0.68 \pm 0.06$ after the first 500 tests, compare to $0.24 \pm 0.05$ for \textit{GA}, $0.17 \pm 0.04$ for $BF$, and only $0.04 \pm 0.02$ for \textit{SA}; by the full 3,900 test budget these averages rise to 0.87, 0.46, 0.39, and 0.11 respectively. While the ensemble-based method consistently outperforms the other approaches in most configurations, its advantage over \textit{GA} in this instance is less pronounced.  
Specifically, the comparison yields a one-sided Wilcoxon $p$-value of 0.048, with a 95\% confidence interval of [0.03, 0.41].  
These results indicate only marginal evidence of superiority, rather than a substantial widening of performance~\cite{Wasserstein2016}.

This type of visualization provides a macroscopic view of method performance per problem and confirms the robustness of the classifier-based approach.

\begin{figure}[ht]
\centering
\begin{tikzpicture}
\begin{axis}[
    width=11cm,
    height=7cm,
    ymin=0, ymax=0.60,
    xlabel={Number of test-cases},
    ylabel={Average failure probability},
    xticklabel style={font=\small}, % or \footnotesize, \scriptsize, \tiny, \large, etc.
    title={Aggregated Probability vs. Test-Case Size (Best Only)},
    grid=major,
    legend style={at={(1.05,1)}, anchor=north west}
]

% Plot for Ens_best
\addplot+[color=blue, mark=*, mark options={fill=blue},line width=1pt, mark size=3pt] 
coordinates {
(100.0, 0.4300) (300.0, 0.4779) (500.0, 0.5247) (700.0, 0.5290)
(900.0, 0.5321) (1100.0, 0.5319) (1300.0, 0.5365) (1500.0, 0.5340)
(1700.0, 0.5337) (1900.0, 0.5334) (2100.0, 0.5367) (2300.0, 0.5343)
(2500.0, 0.5348) (2700.0, 0.5352) (2900.0, 0.5341) (3100.0, 0.5326)
(3300.0, 0.5337) (3500.0, 0.5337) (3700.0, 0.5311) (3900.0, 0.5304)
};
\addlegendentry{Ens\_best}

% Plot for BF_best
\addplot+[color=red, mark=x, mark options={thick}, line width=1pt, mark size=3pt] 
coordinates {
(100.0, 0.0098) (300.0, 0.0186) (500.0, 0.0341) (700.0, 0.0475)
(900.0, 0.0405) (1100.0, 0.0526) (1300.0, 0.0742) (1500.0, 0.0685)
(1700.0, 0.0846) (1900.0, 0.0996) (2100.0, 0.1020) (2300.0, 0.1107)
(2500.0, 0.1048) (2700.0, 0.1135) (2900.0, 0.1202) (3100.0, 0.1196)
(3300.0, 0.1320) (3500.0, 0.1280) (3700.0, 0.1296) (3900.0, 0.1344)
};
\addlegendentry{BF\_best}

% Plot for SA_best
\addplot+[color=green, mark=square*, mark options={fill=green}, line width=1pt, mark size=3pt] coordinates {
(100.0, 0.0062) (300.0, 0.0127) (500.0, 0.0133) (700.0, 0.0265)
(900.0, 0.0242) (1100.0, 0.0250) (1300.0, 0.0216) (1500.0, 0.0268)
(1700.0, 0.0290) (1900.0, 0.0246) (2100.0, 0.0371) (2300.0, 0.0385)
(2500.0, 0.0383) (2700.0, 0.0339) (2900.0, 0.0367) (3100.0, 0.0411)
(3300.0, 0.0438) (3500.0, 0.0478) (3700.0, 0.0451) (3900.0, 0.0369)
};
\addlegendentry{SA\_best}

% Plot for GA_best
\addplot+[color=orange, mark=triangle*, mark options={fill=orange}, line width=1pt, mark size=3pt] coordinates {
(100.0, 0.0999) (300.0, 0.1164) (500.0, 0.1252) (700.0, 0.1144)
(900.0, 0.1240) (1100.0, 0.1272) (1300.0, 0.1287) (1500.0, 0.1414)
(1700.0, 0.1289) (1900.0, 0.1354) (2100.0, 0.1425) (2300.0, 0.1330)
(2500.0, 0.1477) (2700.0, 0.1500) (2900.0, 0.1551) (3100.0, 0.1422)
(3300.0, 0.1633) (3500.0, 0.1465) (3700.0, 0.1546) (3900.0, 0.1535)
};
\addlegendentry{GA\_best}

\end{axis}
\end{tikzpicture}
\caption{Aggregated performance comparison of four methods across all 17 benchmark concurrency problems. The x-axis shows the number of test-cases used in each evaluation, and the y-axis shows the average fault-triggering probability. For each method, the curve represents the mean of the mean of the best test-case's fault-triggering probability across all problems.}
\label{fig:avg-prob-by-size}
\end{figure}
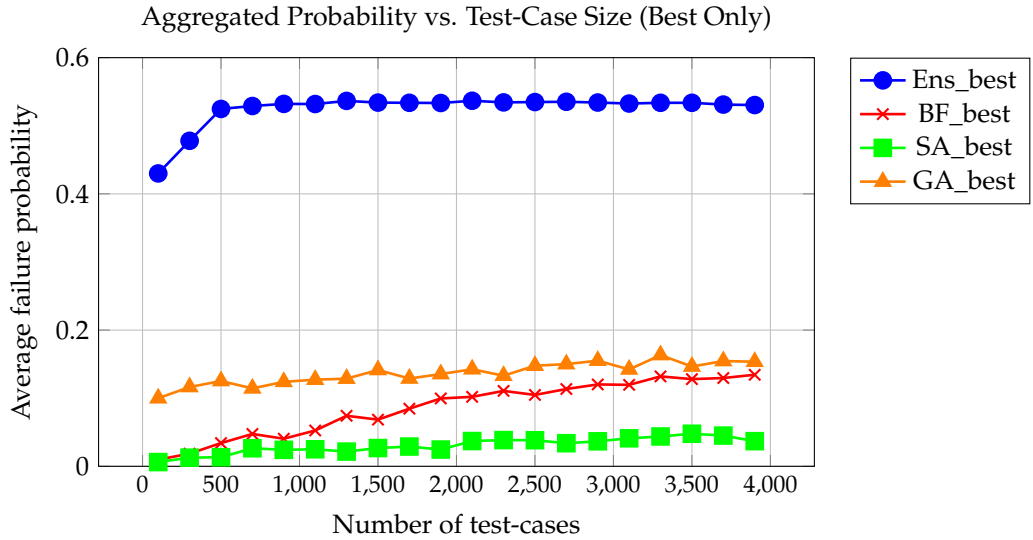

\autoref{fig:avg-prob-by-size} provides a crucial ``bird's-eye view'' of the comparative performance of selected optimization and four searching methods. 
This high-level summary allows one to quickly grasp the overall landscape of method effectiveness without delving into the intricacies of individual experimental variations.
This visualization is generated by processing and aggregating data from all problems' results. The x-axis represents the number of test-cases, while the y-axis indicates the average probability. From these aggregated curves, key insights can be gleaned, such as the convergence behavior of each method as the number of test-cases increases, their relative performance ceilings, and the efficiency with which they approach optimal solutions. 

%-----------------------------------

%xxxxxxxxxxxxxxxxxxxxxxxxxxxxxxxxxxxxxxxxxxxxxxxxxxxxxxxxx

%------------------------------------------------------
\subsection{Per-Problem Bug-Detection Rates}
%------------------------------------------------------

To gain a deeper understanding of method behavior across varying bug difficulties, we divided the 17 benchmark problems into three groups based on their maximum observed bug-detection probabilities: problems with low detectability (maximum probability below 0.2), medium detectability (between 0.2 and 0.6), and high detectability (above 0.6). This classification reflects the intrinsic challenge of each problem and enables structured comparison across problem types. 

For illustrative purposes, we present in this section one representative problem from each group. These examples serve to demonstrate trends that consistently appear across the full suite of benchmarks. In all selected cases, the ensemble method (\textit{Ens}) clearly outperforms the alternatives, both in terms of detection probability and convergence rate. The full set of graphs for all 17 problems is included in the supplementary material.

~\autoref{fig:combined_bugsPP} shows representative cases from a low group (Shared Flag), a moderate group (Atomicity Bypass), and a high group (Race-To-Wait). Here, \textit{Ens} rapidly increases its bug-detection success rate, reaching a median of 50\% after only 1,000 tests. In contrast, the baseline method ($BF$) lags behind at approximately 15\%, while \textit{GA} reaches around 20\%. The \textit{SA} method is the least effective, remaining near zero throughout. This pattern, where \textit{Ens} dominates, \textit{GA} and $BF$ perform moderately, and \textit{SA} struggles, recurs in nearly all problems, regardless of their detectability group.

Quantitatively, \textit{Ens} exceeds the 0.20 success threshold on 9/9 ``low-detectability'' problems, while the next best method (\textit{GA}) manages it on only 3; in the medium tier (max $\in$ [0.2, 0.6]) \textit{Ens} surpasses 0.60 on 5/6 problems vs. 0 for $BF$ and 1 for \textit{GA}; and for high-detectability bugs \textit{Ens} reaches $\ge$ 0.90 on 4/5 problems within 1100 tests, a level $BF$ attains on just one problem.

These results reinforce findings from prior work~\citep{Liu2024TrickCatcher,Ouedraogo2025}, showing that learning-guided techniques not only improve final detection rates, but also significantly reduce the number of test-cases required to reveal faults, particularly in scenarios where failures are elusive or require precise triggering conditions.

%=======================================

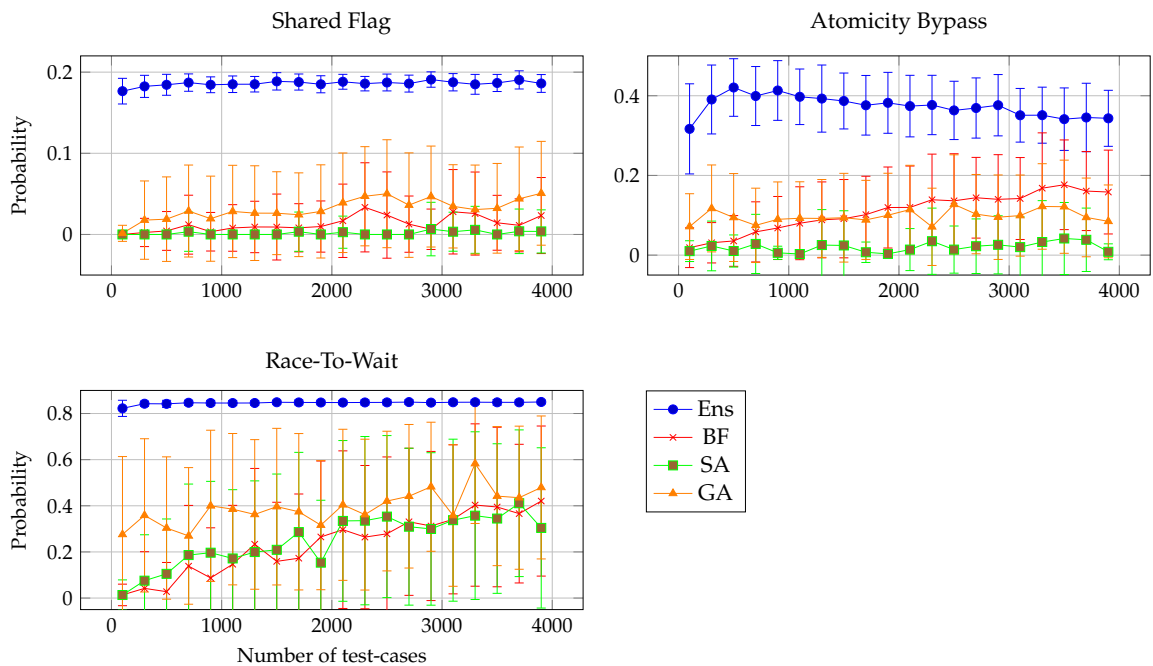
\begin{figure}[H]
\centering
\scalebox{0.85}{
\begin{tikzpicture}
\begin{groupplot}[
    group style={
        group name=bugplots,
        group size=2 by 2, % 1 column, 3 rows
        vertical sep=1.8cm
    },
    width=0.6\columnwidth,
    height=5cm,
    ymin=-0.1, ymax=0.9,
    xtick={0,1000,2000,3000,4000},
    xticklabels={0,1000,2000,3000,4000},
    xticklabel style={font=\small},
    ylabel style={font=\small},
    xlabel style={font=\small},
    tick label style={font=\small},
    grid=major,
    error bars/y dir=both,
    error bars/y explicit,
    legend style={at={(1.125,1)},anchor=north west}    
]

% --- PLOT 1 ---
\nextgroupplot[
    title={Shared Flag},
    ylabel={Probability},
    ymin=-0.05, ymax=0.22,
    xticklabels={0,1000,2000,3000,4000},
]
% ENS + BF + SA + GA for Plot 1
% insert coordinates from your first plot here
\addplot+[mark=*,color=blue,error bars/.cd,y dir=both,y explicit] coordinates {
(100, 0.17656000000000002) +- (0, 0.01580513995444494)
(300, 0.18246000000000007) +- (0, 0.01361303397033707)
(500, 0.1844000000000001) +- (0, 0.012912578901116195)
(700, 0.18712000000000004) +- (0, 0.010695488272526654)
(900, 0.18441999999999997) +- (0, 0.009756400306402744)
(1100, 0.18507999999999997) +- (0, 0.010201720543164098)
(1300, 0.18512) +- (0, 0.009567461864601931)
(1500, 0.18864) +- (0, 0.010693809004298149)
(1700, 0.18780000000000002) +- (0, 0.009913915184512845)
(1900, 0.18504) +- (0, 0.01050356596395642)
(2100, 0.18811999999999995) +- (0, 0.009168802472270317)
(2300, 0.1858) +- (0, 0.008909980645435817)
(2500, 0.18724) +- (0, 0.010385547257450362)
(2700, 0.18587999999999993) +- (0, 0.010434714433376818)
(2900, 0.19088000000000002) +- (0, 0.009479602161634062)
(3100, 0.18754) +- (0, 0.0106545190182478)
(3300, 0.18498) +- (0, 0.012234928254295374)
(3500, 0.18666) +- (0, 0.010535866826982886)
(3700, 0.19041999999999992) +- (0, 0.011194185808640454)
(3900, 0.18604000000000007) +- (0, 0.010969270621111676)
};
% \addlegendentry{Ens}
\addplot+[mark=x,color=red,error bars/.cd,y dir=both,y explicit] coordinates {
(100, 0) +- (0, 0.0)
(300, 0.00246) +- (0, 0.01739482681718907)
(500, 0.00434) +- (0, 0.023893137944111204)
(700, 0.01206) +- (0, 0.03630393805590996)
(900, 0.00336) +- (0, 0.023758787847867998)
(1100, 0.008020000000000001) +- (0, 0.028689392194499272)
(1300, 0.00924) +- (0, 0.0316888958856092)
(1500, 0.00916) +- (0, 0.04062137770350253)
(1700, 0.008199999999999999) +- (0, 0.0297032946705532)
(1900, 0.00976) +- (0, 0.031538483133583754)
(2100, 0.01686) +- (0, 0.04519527246362416)
(2300, 0.03342) +- (0, 0.05491109326233913)
(2500, 0.023920000000000004) +- (0, 0.05315661880188561)
(2700, 0.012720000000000002) +- (0, 0.03465105352228243)
(2900, 0.0064) +- (0, 0.024889961914025354)
(3100, 0.027960000000000002) +- (0, 0.05200901020839699)
(3300, 0.02576) +- (0, 0.051167815139230334)
(3500, 0.014400000000000001) +- (0, 0.033818241666114975)
(3700, 0.011260000000000001) +- (0, 0.0326176426530934)
(3900, 0.0233) +- (0, 0.046839826939850046)
};
% \addlegendentry{BF}
\addplot+[mark=square*,color=green,error bars/.cd,y dir=both,y explicit] coordinates {
(100, 0) +- (0, 0.0)
(300, 0) +- (0, 0.0)
(500, 0) +- (0, 0.0)
(700, 0.0034200000000000003) +- (0, 0.024183051916579927)
(900, 0) +- (0, 0.0)
(1100, 0) +- (0, 0.0)
(1300, 0) +- (0, 0.0)
(1500, 0) +- (0, 0.0)
(1700, 0.0034200000000000003) +- (0, 0.024183051916579927)
(1900, 0) +- (0, 0.0)
(2100, 0.0028000000000000004) +- (0, 0.019798989873223333)
(2300, 0) +- (0, 0.0)
(2500, 0) +- (0, 0.0)
(2700, 0) +- (0, 0.0)
(2900, 0.00666) +- (0, 0.03297098910328563)
(3100, 0.0034000000000000002) +- (0, 0.02404163056034262)
(3300, 0.005699999999999999) +- (0, 0.02896179749749166)
(3500, 0) +- (0, 0.0)
(3700, 0.00388) +- (0, 0.027435743110038047)
(3900, 0.00376) +- (0, 0.02658721497261419)
};
% \addlegendentry{SA}
\addplot+[mark=triangle*,color=orange,error bars/.cd,y dir=both,y explicit] coordinates {
(100, 0.0014000000000000002) +- (0, 0.009899494936611667)
(300, 0.01768) +- (0, 0.048244039496488846)
(500, 0.018860000000000002) +- (0, 0.052063936360860714)
(700, 0.02888) +- (0, 0.0566545709379505)
(900, 0.01944) +- (0, 0.05243583445215569)
(1100, 0.02836) +- (0, 0.05685669777284514)
(1300, 0.02628) +- (0, 0.05833927511041051)
(1500, 0.0262) +- (0, 0.05106458495606024)
(1700, 0.0243) +- (0, 0.05159071661468104)
(1900, 0.0284) +- (0, 0.05740724550729371)
(2100, 0.03906) +- (0, 0.061324632997229574)
(2300, 0.04707999999999999) +- (0, 0.06109672491973991)
(2500, 0.050200000000000015) +- (0, 0.06649965474822145)
(2700, 0.03616) +- (0, 0.06436546164723148)
(2900, 0.046720000000000005) +- (0, 0.0621932603475092)
(3100, 0.034640000000000004) +- (0, 0.05130053904319573)
(3300, 0.030660000000000003) +- (0, 0.05485253328847506)
(3500, 0.03224) +- (0, 0.05526136783965995)
(3700, 0.04394) +- (0, 0.06381299528581101)
(3900, 0.05074000000000001) +- (0, 0.0639584718583002)
};
% \addlegendentry{GA}

% --- PLOT 2 ---
\nextgroupplot[
    title={Atomicity Bypass},
    ymin=-0.05, ymax=0.5,
    xticklabels={0,1000,2000,3000,4000},
]
% ENS + BF + SA + GA for Plot 2
\addplot+[mark=*,color=blue,error bars/.cd,y dir=both,y explicit] coordinates {
(100, 0.3168599999999999) +- (0, 0.11307718332075302)
(300, 0.3906000000000001) +- (0, 0.08647000187631416)
(500, 0.42066000000000003) +- (0, 0.07225002224418857)
(700, 0.39927999999999997) +- (0, 0.07401049015552612)
(900, 0.41318) +- (0, 0.07492055384041901)
(1100, 0.39736) +- (0, 0.06970865316824475)
(1300, 0.3928799999999999) +- (0, 0.08421150243795235)
(1500, 0.38681999999999994) +- (0, 0.07015804025373694)
(1700, 0.37608) +- (0, 0.07481919430690682)
(1900, 0.3822399999999998) +- (0, 0.07645661968644482)
(2100, 0.37401999999999996) +- (0, 0.07724503550522961)
(2300, 0.3767999999999999) +- (0, 0.07449092534788235)
(2500, 0.3631999999999999) +- (0, 0.07316712766966237)
(2700, 0.36922000000000005) +- (0, 0.07552625979331508)
(2900, 0.37624) +- (0, 0.07693133965041905)
(3100, 0.35098) +- (0, 0.06734695120593566)
(3300, 0.3512199999999999) +- (0, 0.07030176645964854)
(3500, 0.34122) +- (0, 0.07839806588618017)
(3700, 0.34542) +- (0, 0.08592072520881172)
(3900, 0.34333999999999987) +- (0, 0.07029256935211957)
};
% \addlegendentry{Ens}
\addplot+[mark=x,color=red,error bars/.cd,y dir=both,y explicit] coordinates {
(100, 0.018680000000000002) +- (0, 0.04991766281812971)
(300, 0.031019999999999985) +- (0, 0.05062204488729619)
(500, 0.03559999999999999) +- (0, 0.0636473062417942)
(700, 0.058359999999999995) +- (0, 0.07541551835108783)
(900, 0.06806) +- (0, 0.07911252900077179)
(1100, 0.08002) +- (0, 0.09149338996978561)
(1300, 0.08847999999999995) +- (0, 0.09526338140510605)
(1500, 0.09147999999999999) +- (0, 0.09826956677986347)
(1700, 0.10160000000000002) +- (0, 0.09633805275260268)
(1900, 0.11949999999999995) +- (0, 0.10164589394118213)
(2100, 0.11974000000000004) +- (0, 0.10385032597003063)
(2300, 0.13898) +- (0, 0.11451058056452469)
(2500, 0.13632) +- (0, 0.11814060111855307)
(2700, 0.14400000000000002) +- (0, 0.10127715055475145)
(2900, 0.14) +- (0, 0.11209999180760584)
(3100, 0.14208) +- (0, 0.10281742090140335)
(3300, 0.16800000000000004) +- (0, 0.13886463824567438)
(3500, 0.17656) +- (0, 0.11253319872964834)
(3700, 0.16080000000000003) +- (0, 0.09916096993869948)
(3900, 0.15832000000000002) +- (0, 0.10519999224024176)
};
% \addlegendentry{BF}
\addplot+[mark=square*,color=green,error bars/.cd,y dir=both,y explicit] coordinates {
(100, 0.01) +- (0, 0.026086981662217994)
(300, 0.023060000000000004) +- (0, 0.06236934424892423)
(500, 0.01034) +- (0, 0.04030678780127663)
(700, 0.028140000000000002) +- (0, 0.07423707473564758)
(900, 0.0058200000000000005) +- (0, 0.016563655812578064)
(1100, 0.0027800000000000004) +- (0, 0.00913233769835165)
(1300, 0.02526) +- (0, 0.08900488635978572)
(1500, 0.024259999999999997) +- (0, 0.08709479002201906)
(1700, 0.007200000000000001) +- (0, 0.02564514514265332)
(1900, 0.0031200000000000004) +- (0, 0.0066781397874264474)
(2100, 0.013919999999999998) +- (0, 0.05272835274947881)
(2300, 0.034879999999999994) +- (0, 0.08312141107796375)
(2500, 0.013920000000000002) +- (0, 0.05917419459631411)
(2700, 0.02238) +- (0, 0.06849337521964874)
(2900, 0.025900000000000003) +- (0, 0.07293224420582407)
(3100, 0.02008) +- (0, 0.07577917435091334)
(3300, 0.03288) +- (0, 0.10385893887226785)
(3500, 0.041740000000000006) +- (0, 0.09038778362462169)
(3700, 0.038500000000000006) +- (0, 0.07954058135085922)
(3900, 0.008360000000000001) +- (0, 0.01997320654256188)
};
% \addlegendentry{SA}
\addplot+[mark=triangle*,color=orange,error bars/.cd,y dir=both,y explicit] coordinates {
(100, 0.07159999999999997) +- (0, 0.08248883659784098)
(300, 0.11719999999999996) +- (0, 0.1090800118243111)
(500, 0.09437999999999999) +- (0, 0.11032288233792184)
(700, 0.07471999999999998) +- (0, 0.09330245747797093)
(900, 0.08985999999999997) +- (0, 0.09383050502618748)
(1100, 0.09258) +- (0, 0.09149664653161128)
(1300, 0.09209999999999999) +- (0, 0.09764978029589587)
(1500, 0.09389999999999997) +- (0, 0.11135093449201448)
(1700, 0.08843999999999999) +- (0, 0.09923794534309124)
(1900, 0.10011999999999997) +- (0, 0.10536634632925028)
(2100, 0.11434000000000001) +- (0, 0.11111799391380568)
(2300, 0.07103999999999999) +- (0, 0.0969775820928292)
(2500, 0.12735999999999997) +- (0, 0.12440778241947172)
(2700, 0.10287999999999997) +- (0, 0.09950810447969988)
(2900, 0.09519999999999999) +- (0, 0.10601270620803444)
(3100, 0.09924000000000001) +- (0, 0.10181166285783544)
(3300, 0.12213999999999998) +- (0, 0.10730635335052514)
(3500, 0.12147999999999998) +- (0, 0.11690851822109914)
(3700, 0.09464) +- (0, 0.09876229140760977)
(3900, 0.08476) +- (0, 0.09133802659375127)
};
% \addlegendentry{GA}

% --- PLOT 3 ---
\nextgroupplot[
    title={Race-To-Wait},
    xlabel={Number of test-cases},
    ylabel={Probability},
    ymin=-0.05, ymax=0.9,
    xticklabels={0,1000,2000,3000,4000},
]
% ENS + BF + SA + GA for Plot 3
\addplot+[mark=*,color=blue,error bars/.cd,y dir=both,y explicit] coordinates {
(100, 0.8224000000000001) +- (0, 0.03505272995860693)
(300, 0.8424600000000002) +- (0, 0.011128580597832283)
(500, 0.8418800000000001) +- (0, 0.015311247033285227)
(700, 0.8466200000000002) +- (0, 0.010903828198745398)
(900, 0.8455199999999999) +- (0, 0.011285244113171755)
(1100, 0.84538) +- (0, 0.011193821181931633)
(1300, 0.8455999999999997) +- (0, 0.009800041649224293)
(1500, 0.84874) +- (0, 0.00857573701241605)
(1700, 0.8480200000000002) +- (0, 0.01049098738568162)
(1900, 0.8477199999999996) +- (0, 0.009484767594755856)
(2100, 0.8473800000000001) +- (0, 0.00905513722723636)
(2300, 0.8478999999999998) +- (0, 0.010677556091242636)
(2500, 0.8482000000000001) +- (0, 0.010997216716524429)
(2700, 0.84954) +- (0, 0.009949279534986758)
(2900, 0.8472) +- (0, 0.008378787599208683)
(3100, 0.8487200000000003) +- (0, 0.009733258782679746)
(3300, 0.8489999999999999) +- (0, 0.00978962382800676)
(3500, 0.8484999999999999) +- (0, 0.012242865477576761)
(3700, 0.8483599999999999) +- (0, 0.010017861599184146)
(3900, 0.85004) +- (0, 0.00803807266979455)
};
\addlegendentry{Ens}
\addplot+[mark=x,color=red,error bars/.cd,y dir=both,y explicit] coordinates {
(100, 0.013600000000000001) +- (0, 0.04670882449510248)
(300, 0.04198) +- (0, 0.15891153873415595)
(500, 0.0277) +- (0, 0.1265817877089427)
(700, 0.13836) +- (0, 0.2632822463146533)
(900, 0.08728) +- (0, 0.2172391741272973)
(1100, 0.1467) +- (0, 0.25517710936300136)
(1300, 0.23329999999999998) +- (0, 0.3285781521051214)
(1500, 0.15909999999999996) +- (0, 0.2561918326216525)
(1700, 0.17278000000000002) +- (0, 0.27832885335919605)
(1900, 0.26489999999999997) +- (0, 0.3291190002514753)
(2100, 0.29648) +- (0, 0.34161353268611255)
(2300, 0.2641800000000001) +- (0, 0.310482224671131)
(2500, 0.27848) +- (0, 0.333279207578393)
(2700, 0.33068) +- (0, 0.31894183719011227)
(2900, 0.3123599999999999) +- (0, 0.32322569103791676)
(3100, 0.34132000000000007) +- (0, 0.32309905222860125)
(3300, 0.4033) +- (0, 0.35192817275050625)
(3500, 0.39472) +- (0, 0.34581327036304393)
(3700, 0.366) +- (0, 0.3005284461390893)
(3900, 0.42052) +- (0, 0.32527603599379523)
};
\addlegendentry{BF}
\addplot+[mark=square*,color=green,error bars/.cd,y dir=both,y explicit] coordinates {
(100, 0.01322) +- (0, 0.06554054361342189)
(300, 0.07544000000000001) +- (0, 0.19892672838335757)
(500, 0.105) +- (0, 0.2378197739843329)
(700, 0.18634) +- (0, 0.30783354964284565)
(900, 0.19668) +- (0, 0.3091933177723963)
(1100, 0.17170000000000002) +- (0, 0.2980828024943021)
(1300, 0.19942000000000007) +- (0, 0.30856501911332745)
(1500, 0.20920000000000002) +- (0, 0.32842711116309936)
(1700, 0.28596) +- (0, 0.3456259013347659)
(1900, 0.15347999999999998) +- (0, 0.27050264927594725)
(2100, 0.33440000000000003) +- (0, 0.3483644584603419)
(2300, 0.3355) +- (0, 0.36475399873316916)
(2500, 0.35314) +- (0, 0.35120759661255524)
(2700, 0.3093999999999999) +- (0, 0.3400908042009382)
(2900, 0.29956) +- (0, 0.33046301222524066)
(3100, 0.33756) +- (0, 0.35114228524886587)
(3300, 0.3572399999999999) +- (0, 0.363502775443465)
(3500, 0.34466) +- (0, 0.3241813935412817)
(3700, 0.41102000000000005) +- (0, 0.31796210080923226)
(3900, 0.3041) +- (0, 0.3473931637267516)
};
\addlegendentry{SA}
\addplot+[mark=triangle*,color=orange,error bars/.cd,y dir=both,y explicit] coordinates {
(100, 0.27577999999999997) +- (0, 0.3375984990052715)
(300, 0.35850000000000004) +- (0, 0.3321674798860923)
(500, 0.30363999999999997) +- (0, 0.3084776815265572)
(700, 0.2694) +- (0, 0.2961200804417295)
(900, 0.3998199999999999) +- (0, 0.3277369957257602)
(1100, 0.38526000000000005) +- (0, 0.3280727775158109)
(1300, 0.36248) +- (0, 0.3242698070359034)
(1500, 0.39626000000000006) +- (0, 0.3394303036271336)
(1700, 0.37416) +- (0, 0.33894020663454705)
(1900, 0.3159799999999999) +- (0, 0.27985116233674023)
(2100, 0.40424000000000004) +- (0, 0.3274675593841007)
(2300, 0.36170000000000013) +- (0, 0.32693107617880024)
(2500, 0.42045999999999994) +- (0, 0.3027181763385117)
(2700, 0.44131999999999993) +- (0, 0.3111262910452796)
(2900, 0.4825199999999999) +- (0, 0.2797525320415772)
(3100, 0.35779999999999995) +- (0, 0.30676734520094007)
(3300, 0.58278) +- (0, 0.2594476306718264)
(3500, 0.4419399999999999) +- (0, 0.3014938595509681)
(3700, 0.43501999999999996) +- (0, 0.3101804274144655)
(3900, 0.4794599999999999) +- (0, 0.3098583751539747)
};
\addlegendentry{GA}

\end{groupplot}
\end{tikzpicture}
}
\caption{Bug-detection rates across three benchmark problems with different detectability levels. Based on 50 runs; error bars = SD.}
\label{fig:combined_bugsPP}
\end{figure}

%======================================

\subsection{Top-k Case Effectiveness}

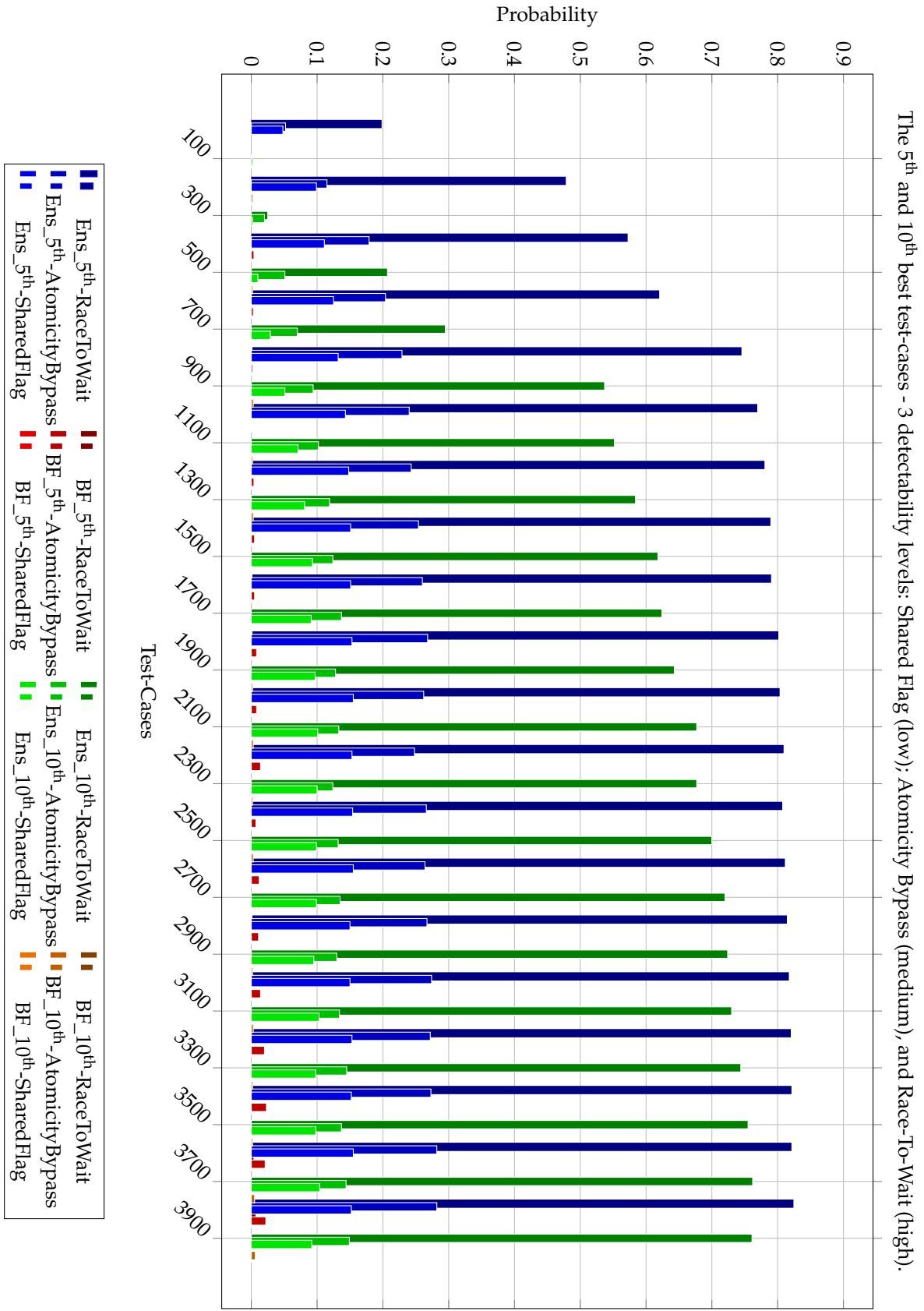
\begin{figure}[]
\centering
\rotatebox{270}{
\begin{tikzpicture}
\begin{axis}[
ybar,
bar width=4pt,
width=21cm,
height=11cm,
symbolic x coords={100-Ens\_5th,100-BF\_5th,100-Ens\_10th,100-BF\_10th,300-Ens\_5th,300-BF\_5th,300-Ens\_10th,300-BF\_10th,500-Ens\_5th,500-BF\_5th,500-Ens\_10th,500-BF\_10th,700-Ens\_5th,700-BF\_5th,700-Ens\_10th,700-BF\_10th,900-Ens\_5th,900-BF\_5th,900-Ens\_10th,900-BF\_10th,1100-Ens\_5th,1100-BF\_5th,1100-Ens\_10th,1100-BF\_10th,1300-Ens\_5th,1300-BF\_5th,1300-Ens\_10th,1300-BF\_10th,1500-Ens\_5th,1500-BF\_5th,1500-Ens\_10th,1500-BF\_10th,1700-Ens\_5th,1700-BF\_5th,1700-Ens\_10th,1700-BF\_10th,1900-Ens\_5th,1900-BF\_5th,1900-Ens\_10th,1900-BF\_10th,2100-Ens\_5th,2100-BF\_5th,2100-Ens\_10th,2100-BF\_10th,2300-Ens\_5th,2300-BF\_5th,2300-Ens\_10th,2300-BF\_10th,2500-Ens\_5th,2500-BF\_5th,2500-Ens\_10th,2500-BF\_10th,2700-Ens\_5th,2700-BF\_5th,2700-Ens\_10th,2700-BF\_10th,2900-Ens\_5th,2900-BF\_5th,2900-Ens\_10th,2900-BF\_10th,3100-Ens\_5th,3100-BF\_5th,3100-Ens\_10th,3100-BF\_10th,3300-Ens\_5th,3300-BF\_5th,3300-Ens\_10th,3300-BF\_10th,3500-Ens\_5th,3500-BF\_5th,3500-Ens\_10th,3500-BF\_10th,3700-Ens\_5th,3700-BF\_5th,3700-Ens\_10th,3700-BF\_10th,3900-Ens\_5th,3900-BF\_5th,3900-Ens\_10th,3900-BF\_10th},
xtick={100-Ens\_10th,300-Ens\_10th,500-Ens\_10th,700-Ens\_10th,900-Ens\_10th,1100-Ens\_10th,1300-Ens\_10th,1500-Ens\_10th,1700-Ens\_10th,1900-Ens\_10th,2100-Ens\_10th,2300-Ens\_10th,2500-Ens\_10th,2700-Ens\_10th,2900-Ens\_10th,3100-Ens\_10th,3300-Ens\_10th,3500-Ens\_10th,3700-Ens\_10th,3900-Ens\_10th},
xticklabels={100,300,500,700,900,1100,1300,1500,1700,1900,2100,2300,2500,2700,2900,3100,3300,3500,3700,3900},
x tick label style={rotate=45,anchor=east},
xlabel={Test-Cases},
ylabel={Probability},
title={The $5^{\text{th}}$ and $10^{\text{th}}$ best test-cases - 3 detectability levels: Shared Flag (low); Atomicity Bypass (medium), and Race-To-Wait (high).},
legend style={at={(0.5,-0.18)},anchor=north,legend columns=4},
ymin=0,ymax=0.9,
enlargelimits=0.05,
scale only axis=true,
grid=major]
\addplot+[
  bar shift=-3pt,
  draw=none,
  fill=black!50!blue
] coordinates {
  (100-Ens\_5th,0.198) (300-Ens\_5th,0.478) (500-Ens\_5th,0.572) (700-Ens\_5th,0.620) (900-Ens\_5th,0.745) (1100-Ens\_5th,0.769) (1300-Ens\_5th,0.780) (1500-Ens\_5th,0.789) (1700-Ens\_5th,0.790) (1900-Ens\_5th,0.801) (2100-Ens\_5th,0.803) (2300-Ens\_5th,0.809) (2500-Ens\_5th,0.807) (2700-Ens\_5th,0.811) (2900-Ens\_5th,0.814) (3100-Ens\_5th,0.817) (3300-Ens\_5th,0.820) (3500-Ens\_5th,0.821) (3700-Ens\_5th,0.821) (3900-Ens\_5th,0.824)
};
\addplot+[
  bar shift=-3pt,
  draw=white,
  fill=black!50!red
] coordinates {
  (100-BF\_5th,nan) (300-BF\_5th,0.000) (500-BF\_5th,0.000) (700-BF\_5th,0.000) (900-BF\_5th,0.000) (1100-BF\_5th,0.000) (1300-BF\_5th,0.000) (1500-BF\_5th,0.000) (1700-BF\_5th,0.000) (1900-BF\_5th,0.000) (2100-BF\_5th,0.000) (2300-BF\_5th,0.000) (2500-BF\_5th,0.000) (2700-BF\_5th,0.000) (2900-BF\_5th,0.000) (3100-BF\_5th,0.001) (3300-BF\_5th,0.000) (3500-BF\_5th,0.000) (3700-BF\_5th,0.004) (3900-BF\_5th,0.007)
};
\addplot+[
  bar shift=0pt,
  draw=white,
  fill=black!50!green
] coordinates {
  (100-Ens\_10th,0.000) (300-Ens\_10th,0.025) (500-Ens\_10th,0.207) (700-Ens\_10th,0.295) (900-Ens\_10th,0.537) (1100-Ens\_10th,0.552) (1300-Ens\_10th,0.584) (1500-Ens\_10th,0.618) (1700-Ens\_10th,0.624) (1900-Ens\_10th,0.643) (2100-Ens\_10th,0.677) (2300-Ens\_10th,0.677) (2500-Ens\_10th,0.700) (2700-Ens\_10th,0.720) (2900-Ens\_10th,0.724) (3100-Ens\_10th,0.730) (3300-Ens\_10th,0.744) (3500-Ens\_10th,0.755) (3700-Ens\_10th,0.762) (3900-Ens\_10th,0.761)
};
\addplot+[
  bar shift=3pt,
  draw=white,
  fill=black!50!orange
] coordinates {
  (100-BF\_10th,nan) (300-BF\_10th,nan) (500-BF\_10th,0.000) (700-BF\_10th,0.000) (900-BF\_10th,0.000) (1100-BF\_10th,0.000) (1300-BF\_10th,0.000) (1500-BF\_10th,0.000) (1700-BF\_10th,0.000) (1900-BF\_10th,0.000) (2100-BF\_10th,0.000) (2300-BF\_10th,0.000) (2500-BF\_10th,0.000) (2700-BF\_10th,0.000) (2900-BF\_10th,0.000) (3100-BF\_10th,0.000) (3300-BF\_10th,0.001) (3500-BF\_10th,0.000) (3700-BF\_10th,0.001) (3900-BF\_10th,0.000)
};
\addplot+[
  bar shift=-1.50pt,
  draw=white,
  fill=black!25!blue
] coordinates {
  (100-Ens\_5th,0.052) (300-Ens\_5th,0.115) (500-Ens\_5th,0.179) (700-Ens\_5th,0.204) (900-Ens\_5th,0.229) (1100-Ens\_5th,0.240) (1300-Ens\_5th,0.243) (1500-Ens\_5th,0.254) (1700-Ens\_5th,0.260) (1900-Ens\_5th,0.268) (2100-Ens\_5th,0.262) (2300-Ens\_5th,0.248) (2500-Ens\_5th,0.266) (2700-Ens\_5th,0.264) (2900-Ens\_5th,0.267) (3100-Ens\_5th,0.274) (3300-Ens\_5th,0.272) (3500-Ens\_5th,0.273) (3700-Ens\_5th,0.282) (3900-Ens\_5th,0.282)
};
\addplot+[
  bar shift=-1.5pt,
  draw=white,
  fill=black!25!red
] coordinates {
  (100-BF\_5th,nan) (300-BF\_5th,0.002) (500-BF\_5th,0.004) (700-BF\_5th,0.003) (900-BF\_5th,0.002) (1100-BF\_5th,0.001) (1300-BF\_5th,0.004) (1500-BF\_5th,0.005) (1700-BF\_5th,0.005) (1900-BF\_5th,0.008) (2100-BF\_5th,0.008) (2300-BF\_5th,0.014) (2500-BF\_5th,0.007) (2700-BF\_5th,0.012) (2900-BF\_5th,0.011) (3100-BF\_5th,0.014) (3300-BF\_5th,0.020) (3500-BF\_5th,0.023) (3700-BF\_5th,0.021) (3900-BF\_5th,0.022)
};
\addplot+[
  bar shift=1.5pt,
  draw=white,
  fill=black!25!green
] coordinates {
  (100-Ens\_10th,0.002) (300-Ens\_10th,0.020) (500-Ens\_10th,0.051) (700-Ens\_10th,0.070) (900-Ens\_10th,0.094) (1100-Ens\_10th,0.102) (1300-Ens\_10th,0.119) (1500-Ens\_10th,0.124) (1700-Ens\_10th,0.137) (1900-Ens\_10th,0.128) (2100-Ens\_10th,0.133) (2300-Ens\_10th,0.124) (2500-Ens\_10th,0.132) (2700-Ens\_10th,0.135) (2900-Ens\_10th,0.130) (3100-Ens\_10th,0.134) (3300-Ens\_10th,0.145) (3500-Ens\_10th,0.137) (3700-Ens\_10th,0.144) (3900-Ens\_10th,0.149)
};
\addplot+[
  bar shift=1.5pt,
  draw=white,
  fill=black!25!orange
] coordinates {
  (100-BF\_10th,nan) (300-BF\_10th,nan) (500-BF\_10th,0.002) (700-BF\_10th,0.001) (900-BF\_10th,0.003) (1100-BF\_10th,0.002) (1300-BF\_10th,0.003) (1500-BF\_10th,0.001) (1700-BF\_10th,0.001) (1900-BF\_10th,0.002) (2100-BF\_10th,0.003) (2300-BF\_10th,0.002) (2500-BF\_10th,0.003) (2700-BF\_10th,0.001) (2900-BF\_10th,0.002) (3100-BF\_10th,0.003) (3300-BF\_10th,0.002) (3500-BF\_10th,0.002) (3700-BF\_10th,0.005) (3900-BF\_10th,0.006)
};
\addplot+[
  bar shift=-0pt,
  draw=white,
  fill=black!10!blue
] coordinates {
  (100-Ens\_5th,0.048) (300-Ens\_5th,0.099) (500-Ens\_5th,0.111) (700-Ens\_5th,0.125) (900-Ens\_5th,0.132) (1100-Ens\_5th,0.143) (1300-Ens\_5th,0.148) (1500-Ens\_5th,0.151) (1700-Ens\_5th,0.151) (1900-Ens\_5th,0.153) (2100-Ens\_5th,0.155) (2300-Ens\_5th,0.153) (2500-Ens\_5th,0.154) (2700-Ens\_5th,0.155) (2900-Ens\_5th,0.150) (3100-Ens\_5th,0.150) (3300-Ens\_5th,0.153) (3500-Ens\_5th,0.152) (3700-Ens\_5th,0.155) (3900-Ens\_5th,0.152)
};
\addplot+[
  bar shift=-0pt,
  draw=white,
  fill=black!10!red
] coordinates {
  (100-BF\_5th,nan) (300-BF\_5th,0.000) (500-BF\_5th,0.000) (700-BF\_5th,0.000) (900-BF\_5th,0.000) (1100-BF\_5th,0.000) (1300-BF\_5th,0.000) (1500-BF\_5th,0.000) (1700-BF\_5th,0.000) (1900-BF\_5th,0.000) (2100-BF\_5th,0.000) (2300-BF\_5th,0.000) (2500-BF\_5th,0.000) (2700-BF\_5th,0.000) (2900-BF\_5th,0.000) (3100-BF\_5th,0.000) (3300-BF\_5th,0.000) (3500-BF\_5th,0.000) (3700-BF\_5th,0.000) (3900-BF\_5th,0.000)
};
\addplot+[
  bar shift=3pt,
  draw=white,
  fill=black!10!green
] coordinates {
  (100-Ens\_10th,0.000) (300-Ens\_10th,0.002) (500-Ens\_10th,0.010) (700-Ens\_10th,0.029) (900-Ens\_10th,0.051) (1100-Ens\_10th,0.071) (1300-Ens\_10th,0.081) (1500-Ens\_10th,0.093) (1700-Ens\_10th,0.091) (1900-Ens\_10th,0.097) (2100-Ens\_10th,0.101) (2300-Ens\_10th,0.100) (2500-Ens\_10th,0.099) (2700-Ens\_10th,0.099) (2900-Ens\_10th,0.095) (3100-Ens\_10th,0.103) (3300-Ens\_10th,0.098) (3500-Ens\_10th,0.098) (3700-Ens\_10th,0.104) (3900-Ens\_10th,0.092)
};
\addplot+[
  bar shift=0pt,
  draw=white,
  fill=black!10!orange
] coordinates {
  (100-BF\_10th,nan) (300-BF\_10th,nan) (500-BF\_10th,0.000) (700-BF\_10th,0.000) (900-BF\_10th,0.000) (1100-BF\_10th,0.000) (1300-BF\_10th,0.000) (1500-BF\_10th,0.000) (1700-BF\_10th,0.000) (1900-BF\_10th,0.000) (2100-BF\_10th,0.000) (2300-BF\_10th,0.000) (2500-BF\_10th,0.000) (2700-BF\_10th,0.000) (2900-BF\_10th,0.000) (3100-BF\_10th,0.000) (3300-BF\_10th,0.000) (3500-BF\_10th,0.000) (3700-BF\_10th,0.000) (3900-BF\_10th,0.000)
};
\legend{Ens\_$5^{\text{th}}$-RaceToWait, BF\_$5^{\text{th}}$-RaceToWait, Ens\_$10^{\text{th}}$-RaceToWait, BF\_$10^{\text{th}}$-RaceToWait, Ens\_$5^{\text{th}}$-AtomicityBypass, BF\_$5^{\text{th}}$-AtomicityBypass, Ens\_$10^{\text{th}}$-AtomicityBypass, BF\_$10^{\text{th}}$-AtomicityBypass, Ens\_$5^{\text{th}}$-SharedFlag, BF\_$5^{\text{th}}$-SharedFlag, Ens\_$10^{\text{th}}$-SharedFlag, BF\_$10^{\text{th}}$-SharedFlag}
\end{axis}
\end{tikzpicture}
}
\caption{The $5^{\text{th}}$ and $10^{\text{th}}$ best test-cases probability.  In three detectability levels of problems that cover three ranges of probability: Shared Flag (low); Atomicity Bypass (medium), and Race-To-Wait (high). Each bar is one experiment and based on 50 independent runs. The X axis is all 20 test-cases, and for each problem, 2 chosen methods ($Ens \& BF$) and 3 (out of 17) problems are shown. The y-axis is the maximum probability for the best test-case.}
\label{fig:5th10thcombinedAll}
\end{figure}

This section compares the performance of \textit{Ens} compare to the $BF$ method when selecting the top-5 and top-10 best test-cases from a larger candidate set. As previously explained, it is often necessary to generate multiple test scenarios for different phases of the process (development, debugging, testing, and validation). Therefore, we demonstrate the ability to generate either the $5^{\text{th}}$ or the $10^{\text{th}}$ test-case according to the two main methods: \textit{Ens} and $BF$. These results, shown in \autoref{fig:5th10thcombinedAll} for $5^{\text{th}}$-best and $10^{\text{th}}$-best for low detectability (Shared Flag), for medium detectability (Atomicity Bypass); and for high detectability (Race-To-Wait), demonstrate how prioritizing test-cases by a classifier model yields a higher likelihood of bug exposure.

%-----------------------------------------------

% --- Combined Graph ---

%---------------------------------------------
We note that for most problems, especially those in the medium-to-high difficulty range (true probability of failure between 0.2 and 0.6), the classifier-based method (\textit{Ens}) consistently outperforms all baselines. On average, the probability of detecting a fault in the top-1 test-case rises from 22\% with $BF$ to 45\% with \textit{Ens}, a relative improvement of more than 100\%. In the top-10 ranking, the average success rate jumps from 40\% ($BF$) to 72\% (\textit{Ens}), with 9 out of 10 problem instances showing a statistically significant advantage (Wilcoxon one-sided test, $p < 0.01$).

These graphs support the hypothesis that even partial ranking from learned models can significantly improve fault detection.

Averaging over the entire benchmark, the $5^{\text{th}}$ best test-case chosen by \textit{Ens} triggers the bug 31\% of the time, vs. 11\% for $BF$; for the $10^{\text{th}}$ best candidate, the rates are 24\% vs. 6\% (both differences significant at $p < 0.001$).

\subsection{Pairwise Statistical Significance Analysis}
\label{sec:statistical-analysis}

This section presents a detailed pairwise statistical comparison between the evaluated methods using one-sided Wilcoxon signed-rank tests, in accordance with contemporary best practices for nonparametric analysis~\cite{Benavoli2015}. \autoref{tab:wilcoxon-best} displays the results across all 17 benchmark problems, using the best-case test input identified for each method. Each row corresponds to a benchmark problem, and each column reports the outcome of a directional hypothesis comparing a pair of methods \textit{Ens}, $BF$, \textit{SA}, and \textit{GA} where aech cell shows the $p$-value for the hypothesis that the method in the row significantly outperforms the method in the column.

Green cells indicate statistically significant superiority ($p < 0.05$), gray cells indicate nonsignificant differences ($p \geq 0.05$), and red cells represent reversed directions. In total, the table comprises 68 directional pairwise comparisons (17 problems $\times$ 4 method pairs). The \textit{Ens} method shows a particularly strong trend: it significantly outperforms $BF$ in 15 out of 17 cases, SA in all 17 cases, and \textit{GA} in 16 cases. This consistency reflects both strong absolute performance and low variability. In contrast, $BF$ significantly outperforms SA in 15 problems but offers a limited advantage over \textit{GA}, outperforming it significantly in only two problems. \textit{GA} and SA, on the other hand, do not significantly outperform any other method in any problem, indicating weaker and less consistent behavior.

Only 14 of the 102 comparisons are statistically inconclusive (gray cells), highlighting that the majority of results are directional and meaningful. This overall structure reveals a clear performance hierarchy: \textit{Ens} consistently outperforms all others, $BF$ performs reliably better than SA, while SA and \textit{GA} rarely, if ever, demonstrate statistical superiority. These patterns underline the robustness of the \textit{Ens} approach across diverse concurrency bug types and failure modes. The statistical evidence supports its adoption as a dominant strategy for test amplification in multithreaded programs.

%5555555555555555555555555555555555555555555

\begin{table}[ht]
\small
\centering
\begin{tabular}{|l|c|c|c|c|c|c|}
\hline
\rowcolor{gray!25}
\textbf{Problem} & \textbf{Ens}$\rightarrow$\textbf{GA} & \textbf{Ens}$\rightarrow$\textbf{BF} & \textbf{Ens}$\rightarrow$\textbf{SA} & \textbf{GA}$\rightarrow$\textbf{BF} & \textbf{GA}$\rightarrow$\textbf{SA} & \textbf{BF}$\rightarrow$\textbf{SA} \\ \hline \hline
AtomicityBypass & \colorbox{green!25}{0.002} & \colorbox{green!25}{<0.001} & \colorbox{green!25}{<0.001} & \colorbox{gray!25}{0.566} & \colorbox{green!25}{<0.001} & \colorbox{green!25}{<0.001} \\ \hline
BrokenBarrier & \colorbox{green!25}{<0.001} & \colorbox{green!25}{<0.001} & \colorbox{green!25}{<0.001} & \colorbox{green!25}{<0.001} & \colorbox{green!25}{<0.001} & \colorbox{green!25}{<0.001} \\ \hline
BrokenPeterson & \colorbox{green!25}{0.002} & \colorbox{green!25}{<0.001} & \colorbox{green!25}{<0.001} & \colorbox{green!25}{<0.001} & \colorbox{green!25}{<0.001} & \colorbox{green!25}{<0.001} \\ \hline
DelayedWrite & \colorbox{green!25}{0.003} & \colorbox{green!25}{<0.001} & \colorbox{green!25}{<0.001} & \colorbox{green!25}{<0.001} & \colorbox{green!25}{<0.001} & \colorbox{green!25}{<0.001} \\ \hline
FlagedDeadlock & \colorbox{green!25}{0.003} & \colorbox{green!25}{0.002} & \colorbox{green!25}{<0.001} & \colorbox{green!25}{<0.001} & \colorbox{green!25}{<0.001} & \colorbox{green!25}{<0.001} \\ \hline
IfNotWhile & \colorbox{green!25}{0.003} & \colorbox{green!25}{0.003} & \colorbox{green!25}{0.001} & \colorbox{green!25}{<0.001} & \colorbox{green!25}{<0.001} & \colorbox{green!25}{<0.001} \\ \hline
LockOrderInversion & \colorbox{red!25}{0.984} & \colorbox{gray!25}{0.434} & \colorbox{green!25}{0.003} & \colorbox{gray!25}{0.054} & \colorbox{green!25}{<0.001} & \colorbox{green!25}{<0.001} \\ \hline
LostSignal & \colorbox{green!25}{<0.001} & \colorbox{green!25}{<0.001} & \colorbox{green!25}{<0.001} & \colorbox{gray!25}{0.174} & \colorbox{green!25}{<0.001} & \colorbox{green!25}{<0.001} \\ \hline
PartialLock & \colorbox{gray!25}{0.295} & \colorbox{gray!25}{0.214} & \colorbox{green!25}{0.003} & \colorbox{gray!25}{0.130} & \colorbox{green!25}{<0.001} & \colorbox{green!25}{<0.001} \\ \hline
PhantomPermit & \colorbox{green!25}{0.003} & \colorbox{green!25}{0.003} & \colorbox{green!25}{0.003} & \colorbox{gray!25}{0.127} & \colorbox{green!25}{0.003} & \colorbox{green!25}{0.011} \\ \hline
RaceToWait & \colorbox{green!25}{0.007} & \colorbox{green!25}{0.003} & \colorbox{green!25}{0.003} & \colorbox{green!25}{<0.001} & \colorbox{green!25}{<0.001} & \colorbox{red!25}{0.996} \\ \hline
RacyIncrement & \colorbox{green!25}{0.003} & \colorbox{green!25}{<0.001} & \colorbox{green!25}{<0.001} & \colorbox{green!25}{<0.001} & \colorbox{green!25}{<0.001} & \colorbox{green!25}{<0.001} \\ \hline
SemaphoreLeak & \colorbox{green!25}{<0.001} & \colorbox{green!25}{<0.001} & \colorbox{green!25}{<0.001} & \colorbox{green!25}{<0.001} & \colorbox{green!25}{<0.001} & \colorbox{green!25}{<0.001} \\ \hline
SharedCounter & \colorbox{green!25}{0.003} & \colorbox{green!25}{0.003} & \colorbox{green!25}{0.001} & \colorbox{green!25}{<0.001} & \colorbox{green!25}{<0.001} & \colorbox{green!25}{<0.001} \\ \hline
SharedFlag & \colorbox{green!25}{0.003} & \colorbox{green!25}{0.003} & \colorbox{green!25}{<0.001} & \colorbox{green!25}{<0.001} & \colorbox{green!25}{<0.001} & \colorbox{green!25}{<0.001} \\ \hline
SignalThenWait & \colorbox{green!25}{0.002} & \colorbox{green!25}{<0.001} & \colorbox{green!25}{<0.001} & \colorbox{green!25}{<0.001} & \colorbox{green!25}{<0.001} & \colorbox{green!25}{0.014} \\ \hline
SleepingGuard & \colorbox{green!25}{<0.001} & \colorbox{green!25}{<0.001} & \colorbox{green!25}{<0.001} & \colorbox{green!25}{<0.001} & \colorbox{green!25}{<0.001} & \colorbox{green!25}{<0.001} \\ \hline
\end{tabular}
\caption{Wilcoxon one-sided signed-rank test results on \textbf{best} scores. Each cell shows the $p$-value for the hypothesis that the left method performs better than the right (e.g., Ans$\rightarrow$BF). \colorbox{green!25}{Green cells} indicate significant results ($p\le0.05$), \colorbox{gray!25}{gray cells} indicate no significance ($0.05<p<0.95$), and \colorbox{red!25}{red cells} indicate evidence in the opposite direction.} \label{tab:wilcoxon-best}
\end{table}

\subsection{Convergence Analysis Across Methods}
\label{sec:convergence-analysis}

In this analysis, we study convergence patterns by plotting the probability of success as a function of the number of test-cases, aggregated over all 17 benchmark problems. \autoref{fig:OVall} provides a bird's-eye view that captures performance trends at four representative budget levels: 500, 1100, 2100, and 3900 test-cases. For each method, \textit{Ens}, $BF$, \textit{SA}, and \textit{GA}, we plot the best failure-inducing probability obtained per problem, averaged over 50 independent runs.

The ensemble classifier-based method (\textit{Ens}) exhibits remarkably fast and stable convergence. At just 500 test-cases, \textit{Ens} already achieves a mean success probability of 51.8\% across all problems. This value rises to 56.4\% at 1100, 58.7\% at 2100, and reaches 59.8\% at 3900. These gains are not only large in absolute terms but also consistently achieved across a diverse range of problem types. This demonstrates the model's ability to generalize its learned prioritization across different failure patterns.

In contrast, the brute-force approach ($BF$) converges slowly. It begins with a mean success rate of just 3.1\% at 500 test-cases, improving modestly to 6.2\% at 1100, 10.0\% at 2100, and only 13.6\% at 3900 test-cases. This linear and limited improvement confirms the inefficiency of uninformed random exploration.

Simulated Annealing (\textit{SA}) and Genetic Algorithm (\textit{GA}) fall between these extremes. \textit{SA} improves from 1.5\% (at 500 test-cases) to 3.9\% (at 3900), with substantial stagnation between checkpoints, reflecting a limited capacity to escape local minima. \textit{GA} achieves higher starting performance at 500 test-cases (mean 8.1\%) and improves more rapidly than \textit{SA}, reaching 17.3\% at 3900, but still falls far short of \textit{Ens}.

Overall, these convergence patterns reinforce the strength of learning-guided strategies. \textit{Ens} not only achieves the highest final probabilities but also reaches them faster, demonstrating both sample efficiency and consistent generalization. This advantage is particularly valuable in real-world testing scenarios where test execution budgets are constrained and high-probability failure discovery is critical.

%----------------------------------

\subsection{Summary of Key Findings}
\label{sec:summary-findings}

Our evaluation of four amplification methods, Brute-Force ($BF$), Simulated Annealing (\textit{SA}), Genetic Algorithm (\textit{GA}), and Ensemble Classification (\textit{Ens}), across 17 benchmark concurrency problems, led to several key findings that integrate both method-specific behavior and cross-cutting insights:

\begin{description}

\item[\textbf{Learning-based amplification significantly outperforms uninformed approaches.}]
The ensemble classifier (\textit{Ens}) consistently achieved the highest bug-triggering probabilities across nearly all test-case budgets and problems. With just 500 test-cases, \textit{Ens} reached average success probabilities exceeding 0.53, whereas $BF$, \textit{SA}, and \textit{GA} remained below 0.13. At the full budget of 3900 test-cases, \textit{Ens} achieved near-perfect detection (over 0.9 probability) in more than half of the problems, including \textit{LockOrderInversion}, \textit{SignalThenWait}, and \textit{IfNotWhile}.

\item[\textbf{\textit{Ens} converges faster and with fewer test-cases.}]
While $BF$, \textit{SA}, and \textit{GA} showed gradual or erratic improvements, \textit{Ens} rapidly identified failure-inducing cases. For example, in \textit{RacyIncrement}, \textit{Ens} surpassed 0.9 success probability with fewer than 1100 test-cases, while \textit{GA} plateaued at 0.07 and $BF$ at 0.03 even after 3900 cases. This sample efficiency makes \textit{Ens} especially valuable for real-world systems with costly or time-limited testing resources.

\item[\textbf{Traditional search methods offer limited scalability.}]
$BF$ showed minimal improvement over increasing test budgets, with average performance rarely exceeding 0.15 across problems. \textit{SA}'s performance improved modestly but remained inconsistent, failing to trigger bugs in several hard problems like \textit{SharedFlag} and \textit{SemaphoreLeak}. \textit{GA} was more effective than $BF$ and \textit{SA} in moderately complex problems but still lagged behind \textit{Ens} in both speed and final success rates.

\item[\textbf{Problem hardness varies significantly and affects method effectiveness.}]
Some problems were consistently easy (e.g., \textit{SignalThenWait} and \textit{LockOrderInversion}) and triggered by all methods to varying degrees. Others, such as \textit{SharedFlag}, \textit{SemaphoreLeak}, and \textit{BrokenBarrier}, remained elusive, with only \textit{Ens} achieving meaningful success (e.g., 0.49 in \textit{SemaphoreLeak} vs. <0.03 for others). This suggests that learning-based methods are better suited for navigating complex or deceptive search spaces.

\item[\textbf{\textit{Ens} robustness is evident across all tested budgets.}]
The bird's-eye view (\autoref{fig:OVall}) shows that across all 17 problems and at every tested budget (500, 1100, 2100, and 3900), \textit{Ens} consistently led or tied for the highest success rate. Notably, in 13 out of 17 problems, \textit{Ens} reached probabilities above 0.85 with 3900 test-cases, while \textit{GA} exceeded 0.5 in only 7, \textit{SA} in two, and $BF$ in one.

\item[\textbf{Integration of feedback powers \textit{Ens} performance.}]
Unlike the other methods, which rely on sampling or mutation heuristics, \textit{Ens} uses supervised learning to predict and prioritize high-risk inputs. This allows it to generalize from early failures, focusing search efforts efficiently. The result is not only higher probabilities of detecting bugs but also significantly fewer wasted executions.

\item[\textbf{Ablation Study.}]
We conducted ablation studies by removing components from the ensemble classifier and modifying its sampling heuristics. Specifically, we evaluated simplified variants of our pipeline, such as omitting SMOTE or disabling passthrough to the meta-learner. These reduced versions consistently underperformed relative to the full classifier configuration we present in the paper. In several cases, the simplified ensemble-based methods even performed worse than the brute-force baseline, highlighting the importance of each pipeline component in achieving effective bug amplification.

\end{description}

Our findings support the superiority of learning-guided search for amplifying concurrency bugs. \textit{Ens} is not only more effective in absolute terms but also more efficient, scalable, and robust across problem domains and budgets. These characteristics make it a promising default choice for future automated bug-amplification tools.

%------------------------------------------------------

\section{Related Work}
\subsection{Concurrency Bug Debugging Methods}

Over the past ten years, concurrent systems bug hunting has evolved significantly, driven by the growing complexity of multithreaded software and the critical need to detect concurrency bugs, such as data races, deadlocks, and atomicity violations.

A survey of academic papers from sources like IEEE Xplore, ACM Digital Library, and SpringerLink reveals three dominant methodological categories: static analysis, dynamic analysis, and model checking, each encompassing diverse techniques with unique trade-offs, industrial applications, and ongoing refinements.

\textbf{Static analysis:}
Techniques that scrutinize code without execution include abstract interpretation, data-flow analysis, type systems, symbolic execution, and machine learning-based bug prediction. Abstract interpretation~\cite{MightVanHorn2011} models program semantics to detect bugs across all paths, offering early detection but often producing false-positives due to over-approximation. Data-flow analysis~\cite{Bora2021LLVMHPC} tracks dependencies and works well in structured parallelism (e.g., OpenMP), though its generalization to unstructured concurrency remains limited. Type systems, such as Rust's ownership model~\cite{MatsakisKlock2014}, prevent bugs at compile-time with minimal runtime cost, though they require full language adoption. Symbolic execution~\cite{Sen2005DART} can uncover deep concurrency bugs through path exploration but suffers from path explosion. Machine learning approaches~\cite{Tehrani2019DeepRace} learn patterns from code to predict concurrency bugs but depend heavily on the availability of labeled data. Tools like Coverity leverage static analysis in the industry, though concurrency-specific precision remains a challenge.

\textbf{Dynamic analysis:}
This category executes programs to observe runtime-behavior and includes methods like thread-aware fuzzing, runtime monitoring, and record-and-replay. Thread-aware fuzzing~\cite{Chen2020MUZZ} explores interleavings to expose real bugs but may suffer from incomplete coverage. Runtime monitoring~\cite{Roemer2020SmartTrack} provides precise race detection at the cost of performance overhead. Record-and-replay~\cite{OCallahan2017RR} facilitates debugging by reproducing execution paths, albeit with recording overhead. Tools like ThreadSanitizer are widely used due to their balance of effectiveness and performance.

\textbf{Model checking:}
This technique provides formal verification by exhaustively exploring program state-spaces. Explicit-state model checking~\cite{Holzmann1997SPIN} can prove correctness but is vulnerable to state explosion. Bounded model checking~\cite{Clarke2001BMC} uses SAT/SMT solvers to explore execution within depth bounds, trading completeness for scalability. Abstraction-based techniques~\cite{Clarke2000Abstraction} simplify systems but risk imprecision. Compositional approaches~\cite{NamjoshiTrefler2016} decompose systems for modular checking, though assumptions can break down. Statistical model checking~\cite{Legay2019SMC} approximates correctness via sampling and is used in domains like embedded systems and aerospace, where formal guarantees are difficult to obtain.

Hybrid approaches have emerged to balance strengths and weaknesses, e.g., KRACE~\cite{Xu2020KRACE} employs thread-scheduling perturbation and fuzzing to detect data races in kernel file systems. Benchmarks such as the Linux kernel and SPEC CPU continue to reveal challenges: static methods must reduce false-positives, dynamic tools need improved coverage, and model checking must scale better. Future directions involve tighter integration of these methods and greater automation to support concurrency bug detection at scale.

\subsection{Concurrency Bug Datasets}

The study of concurrency bugs has led to the development of a wide range of datasets, each designed to capture specific aspects of concurrent programming behavior. These datasets can be grouped into four broad categories: general-purpose concurrency bug datasets, language-specific datasets, smart contract datasets, and fuzzing-based datasets. Below, we summarize key datasets from each category, highlighting their structure, scope, and contributions to academic research.

\textbf{General-Purpose Concurrency Bug Datasets:}
Early work in concurrency bug research focused on real-world software systems. ~\cite{Lu2008Learning} compiled 105 concurrency bugs from widely used applications such as MySQL and Apache. The dataset revealed common bug patterns and has influenced numerous studies in static and dynamic analysis. CHESS~\cite{Musuvathi2008Heisenbugs}, developed by Microsoft Research, explores all thread interleaving to find concurrency bugs. RACEBENCH~\cite{Tian2011RaceBench} is a benchmark suite containing 29 multithreaded programs with known races, offering a standardized environment for testing race detectors. DETECT~\cite{Zhang2011ConSeq} uses dynamic analysis and communication graphs to identify concurrency bugs.

\textbf{Language-Specific Datasets:}
With the growing demand for language-aware tools, several datasets were created targeting Java and Go. For Java, JaConTeBe~\cite{Lin2015JaConTeBe} includes 47 confirmed bugs from 8 Java projects. Defects4J~\cite{Just2014Defects4J} is a curated repository of real-world Java bugs, used extensively in software testing and repair. Bears~\cite{Madeiral2019Bears} collects bugs from CI pipelines to support automated program repair. ManySStuBs4J~\cite{Karampatsis2020ManySStuBs4J} offers over 500k single-statement bugs, indirectly supporting concurrency research. For Go, the Go Concurrency Bug Collection~\cite{Tu2019GoConcurrency} contains 171 bugs from six Go applications~\footnote{\url{[https://github.com/system-pclub/go-concurrency-bugs}}. GoBench~\cite{Yuan2021GoBench} expands this effort with 82 real bugs and 103 bug kernels.

\textbf{Smart Contract Datasets:}
With the rise of blockchain applications, concurrency issues in smart contracts gained prominence. ConFuzzius~\cite{Torres2021ConFuzzius} combines evolutionary fuzzing and symbolic execution to detect concurrency-related bugs in Ethereum smart contracts, building a dataset of known vulnerabilities.

\textbf{Fuzzing-Based Datasets:}
Grey-box fuzzing has proven valuable for stress-testing concurrent applications. MUZZ~\cite{Chen2020MUZZ} presents a thread-aware fuzzing method for multithreaded programs, featuring a dataset of real-world apps annotated with concurrency bugs.

These datasets continue to support advances in concurrency research, enabling reproducibility, benchmarking, and tool evaluation across diverse programming environments.

\section{Detailed Description of the Benchmark Problems}     
\label{sec:benchmark-detailed}

This section details the benchmark problems. For every problem, we document (i) the scenario, (ii) its observable effect, (iii) the underlying root cause, and (iv) a concise insight: 

%=============================

\subsection{Atomicity Bypass: Unexpected Data from Lock Misuse}

\label{sec:prob17}

\begin{description}
\item[Description:] Simulates two threads updating a shared counter under the false assumption that a critical section is properly protected. Each thread acquires a mutex, reads the counter, but then mistakenly releases the mutex before performing the update. As a result, both threads read the same value (e.g., 0), and both write back 1, overwriting each other's increment. The final result is data corruption: the counter appears to have only been incremented once.

\item[Effect:] A clearly unexpected data outcome, where both threads read the same initial value of the counter and write back identical updates, resulting in a lost increment. This leads to data corruption, as the counter reflects only one update instead of two, violating correctness expectations.

\item[Root Cause:] Misuse of concurrency primitives: The locking discipline was violated by releasing the mutex too early.

\item[Insight:] This demonstrates that simply using synchronization tools is insufficient - they must be used correctly and consistently to protect shared operations.

\item[Pseudo Code:]{\phantom{.}}

\begin{minipage}{0.48\textwidth}
\begin{algorithm}[H]
\caption*{Thread 0}
\begin{algorithmic}[1]
\While{mutex == 1}
    \State wait()
\EndWhile
\State mutex $\gets$ 1
\State local $\gets$ counter
\State mutex $\gets$ 0 \Comment{BUG: unlock before update}
\State counter $\gets$ local + 1
\end{algorithmic}
\end{algorithm}
\end{minipage}
\hspace{0.5cm}
\begin{minipage}{0.35\textwidth}
\begin{algorithm}[H]
\caption*{Thread 1}
\begin{algorithmic}[1]
\While{mutex == 1}
    \State wait()
\EndWhile
\State mutex $\gets$ 1
\State local $\gets$ counter
\State mutex $\gets$ 0
\State counter $\gets$ local + 1
\end{algorithmic}
\end{algorithm}
\end{minipage}
\end{description}

\subsection{Broken Barrier: Deadlock from Barrier Misuse with Incorrect Participant Count}

\label{sec:prob6}

\begin{description}
\item[Description:] Three threads increment a shared variable and call \texttt{SignalAndWait()} on a barrier that is configured for only two participants. One thread calls \texttt{SignalAndWait()} twice before resetting the barrier, violating the expected usage pattern.

\item[Effect:] This misconfiguration can lead to deadlock, as some threads may wait indefinitely for signals that never arrive. It may also cause assertion failures if the synchronization logic assumes a specific number of participants.

\item[Root Cause:] A misuse of primitives, where the barrier is used in a way that contradicts its intended configuration.

\item[Insight:] This problem illustrates the importance of synchronization primitives being correctly configured for the actual number of participating threads. Misuse of barriers can lead to subtle and difficult-to-diagnose concurrency failures.

\item[Pseudo Code:]{\phantom{.}}

\begin{minipage}{0.42\textwidth}
\begin{algorithm}[H]
\caption*{Thread 0}
\begin{algorithmic}[1]
\While{true}
  \State Increment(ref fireballCharge)
  \State barrier.SignalAndWait()
  \If{fireballCharge < 2}
    \State Debug.Assert(false)
  \EndIf
  \State fireball()
\EndWhile
\end{algorithmic}
\end{algorithm}
\end{minipage}
\hspace{0.5cm}
\begin{minipage}{0.42\textwidth}
\begin{algorithm}[H]
\caption*{Thread 1}
\begin{algorithmic}[1]
\While{true}
  \State Increment(ref fireballCharge)
  \State barrier.SignalAndWait()
\EndWhile
\end{algorithmic}
\end{algorithm}
\end{minipage}

\begin{minipage}{0.65\textwidth}
\begin{algorithm}[H]
\caption*{Thread 2}
\begin{algorithmic}[1]
\While{true}
  \State Increment(ref fireballCharge)
  \State barrier.SignalAndWait()
  \State barrier.SignalAndWait()
  \State fireballCharge $\gets 0$ \Comment{BUG: reset can occur too early}
\EndWhile
\end{algorithmic}
\end{algorithm}
\end{minipage}
\end{description}

\subsection{Broken Peterson: Mutual Exclusion Violation in Generalized Peterson's Algorithm}

\label{sec:prob11}

\begin{description}
\item[Description:] This problem involves a generalized version of Peterson's algorithm for four processes. The implementation uses arrays to track process levels and a \texttt{last\_to\_enter} array to manage entry ordering. However, a critical assignment to \texttt{last\_to\_enter[level]} is omitted, breaking the algorithm's tie-breaking logic.

\item[Effect:] Multiple processes may enter the critical section concurrently, leading to a concurrent access.

\item[Root Cause:] A missing or weak guard in the synchronization protocol, specifically, a missing update in the entry coordination mechanism.

\item[Insight:] This example highlights how even small implementation errors in well-established algorithms can undermine their correctness. It underscores the need for rigorous validation of synchronization logic, especially in generalized or modified versions of classic algorithms.

\item[Pseudo Code:]{\phantom{.}}

\begin{minipage}{0.85\textwidth}
\begin{algorithm}[H]
\caption*{General Peterson Algorithm (Process $i$)}
\begin{algorithmic}[1]
\While{true}
\For{$\ell = i$ to $n - 2$}
\State last\_to\_enter[$\ell$] $\gets i$ \Comment{{Bug: wrong order}}
\State levels[$i$] $\gets \ell$
\While{\textbf{exists} $k \ne i$ such that levels[$k$] $\ge$ $\ell$ and last\_to\_enter[$\ell$] = $i$}
\State \textbf{wait}
\EndWhile
\EndFor
\State critical\_section()
\State levels[$i$] $\gets -1$
\State remainder\_section()
\EndWhile
\end{algorithmic}
\end{algorithm}
\end{minipage}
\end{description}

\subsection{Delayed Write – Assertion Failure from Non-Atomic Test-and-Set Simulation}

\label{sec:prob10}

\begin{description}
\item[Description:] A simulation models a \texttt{test-and-set} operation where one thread sets a shared variable \texttt{x} to a target value. However, another thread may interleave and modify \texttt{x} during a context switch, violating the assumption that \texttt{x} remains unchanged after being set.

\item[Effect:] A concurrent access and unexpected data, often manifesting as an assertion failure when the invariant \texttt{x == target} is violated.

\item[Root Cause:] An incorrect command ordering stemming from the test-and-set logic. The thread reads and later writes to \texttt{x}, but a context switch between these steps allows another thread to intervene and modify the variable, violating expected execution order.

\item[Insight:] This case illustrates how concurrency bugs can emerge even in simulated atomic operations if the underlying memory operations are not properly synchronized. It emphasizes the importance of true atomicity in synchronization primitives.

\item[Pseudo Code:]{\phantom{.}}

\begin{minipage}{0.35\textwidth}
\begin{algorithm}[H]
\caption*{Thread 0}
\begin{algorithmic}[1]
\State global x
\State x=TARGET
\State if x != TARGET:
\State\hspace{0.5cm}assert (x!=TARGET)

\end{algorithmic}
\end{algorithm}
\end{minipage}
\hspace{0.5cm}
\begin{minipage}{0.25\textwidth}
\begin{algorithm}[H]
\caption*{Thread 1}
\begin{algorithmic}[1]
\State global x
\State x = $3$
\end{algorithmic}
\end{algorithm}
\end{minipage}
\end{description}

\subsection{Flagged Deadlock: Deadlock Risk from Complex Locking}

\label{sec:prob5}

\begin{description}
\item[Description:] Involves two threads using a combination of locking strategies, including recursive locks, try-locks, and conditional logic based on shared flags. The complexity of the locking protocol introduces multiple paths for acquiring locks, some of which may conflict or fail to release locks properly.

\item[Effect:] A heightened risk of deadlock, as threads may become stuck waiting for locks that are never released or acquired in inconsistent orders.

\item[Root Cause:] A combination of misuse of primitives and non-cooperative scheduling, exacerbated by the use of active waiting (spin locks) instead of blocking synchronization.

\item[Insight:] This case highlights the dangers of over-engineering synchronization logic. Complex locking schemes, especially those involving conditional paths and re-entrant locks, are prone to subtle bugs and should be avoided in favor of simpler, more predictable designs.

\item[Pseudo Code:]{\phantom{.}}

\noindent
\begin{minipage}{0.41\textwidth}
\begin{algorithm}[H]
\caption*{Thread 0}
\begin{algorithmic}[1]
\While{true}
  \If{Monitor.TryEnter(mutex)}
    \State Monitor.Enter(mutex3)
    \State Monitor.Enter(mutex)
    \State critical\_section()
    \State Monitor.Exit(mutex)
    \State Monitor.Enter(mutex2)
    \State flag $\gets$ false
    \State Monitor.Exit(mutex2)
    \State Monitor.Exit(mutex3)
  \Else
    \State Monitor.Enter(mutex2)
    \State flag $\gets$ true
    \State Monitor.Exit(mutex2)
  \EndIf
\EndWhile
\end{algorithmic}
\end{algorithm}
\end{minipage}

% \hspace{0.25cm}
\begin{minipage}{0.60\textwidth}
\begin{algorithm}[H]
\caption*{Thread 1}
\begin{algorithmic}[1]
\While{true}
  \If{flag}
    \State Monitor.Enter(mutex2)
    \State Monitor.Enter(mutex) \Comment{BUG: mutex is held}
    \State flag $\gets$ false
    \State critical\_section()
    \State Monitor.Exit(mutex)
    \State Monitor.Enter(mutex2) \Comment{BUG:already held it}
  \Else
    \State Monitor.Enter(mutex)
    \State flag $\gets$ false
    \State Monitor.Exit(mutex)
  \EndIf
\EndWhile
\end{algorithmic}
\end{algorithm}
\end{minipage}
\end{description}

\subsection{If-Not-While: Deadlock and Missed Signals from Condition Variable Misuse}

\label{sec:prob8}

\begin{description}
\item[Description:] Two consumer threads wait on a shared queue using \texttt{Monitor.Wait(mutex)} when the queue is empty. A producer thread enqueues data and signals all waiting consumers using \texttt{Monitor.PulseAll(mutex)}. However, the consumers guard the wait with an \texttt{if} statement rather than a \texttt{while} loop, failing to re-check the condition upon waking.

\item[Effect:] This leads to two possible effects: deadlock, if a consumer misses a signal and waits indefinitely, or unexpected data loss, if a consumer proceeds without the queue being properly populated.

\item[Root Cause:] A race condition caused by a weak guard; the failure to revalidate the condition after waking allows incorrect assumptions about the system state.

\item[Insight:] This problem reinforces the importance of using guarded waits with \texttt{while} loops when working with condition variables, ensuring that threads only proceed when the condition they depend on is truly satisfied.

\item[Pseudo Code:]{\phantom{.}}

\begin{minipage}{0.55\textwidth}
\begin{algorithm}[H]
\caption*{Thread 0}
\begin{algorithmic}[1]
\While{true}
  \State Monitor.Enter(mutex)
  \If{queue.Count == 0}
    \State Monitor.Wait(mutex) \Comment{release \& wait}
  \EndIf
  \State queue.Dequeue()
  \State Monitor.Exit(mutex)
\EndWhile
\end{algorithmic}
\end{algorithm}
\end{minipage}

\begin{minipage}{0.55\textwidth}
\begin{algorithm}[H]
\caption*{Thread 1}
\begin{algorithmic}[1]
\While{true}
  \State Monitor.Enter(mutex)
  \If{queue.Count == 0}
    \State Monitor.Wait(mutex) \Comment{release \& wait}
  \EndIf
  \State queue.Dequeue()
  \State Monitor.Exit(mutex)
\EndWhile
\end{algorithmic}
\end{algorithm}
\end{minipage}
\hspace{0.5cm}
\begin{minipage}{0.35\textwidth}
\begin{algorithm}[H]
\caption*{Thread 2}
\begin{algorithmic}[1]
\While{true}
  \State Monitor.Enter(mutex)
  \State queue.Enqueue(42)
  \State Monitor.PulseAll(mutex)
  \State Monitor.Exit(mutex)
\EndWhile
\end{algorithmic}
\end{algorithm}
\end{minipage}
\end{description}

\subsection{Lock Order Inversion: Deadlock from Inconsistent Lock Acquisition Order}

\label{sec:prob1}
\begin{description}
\item [Description:] In this classic concurrency scenario, two threads attempt to acquire two shared locks but do so in opposite orders. Thread 0 first locks \texttt{mutex1} and then attempts to acquire \texttt{mutex2}, while Thread 1 begins by locking \texttt{mutex2} and then proceeds to request \texttt{mutex1}. This inversion in lock acquisition order creates a circular wait condition: each thread holds one lock and waits indefinitely for the other to release the second, which never happens.

\item[Effect:] A deadlock, where both threads are permanently blocked, unable to make progress.

\item[Root Cause:] An incorrect order, a well-known concurrency design flaw where multiple threads acquire shared resources in inconsistent sequences. When such errors occur, they can easily lead to circular dependencies, especially in systems that lack a global lock acquisition policy.

\item[Insight:] This problem exemplifies the dangers of uncoordinated locking strategies in multithreaded environments. It highlights the importance of enforcing a consistent global order for acquiring multiple locks, a practice that can prevent deadlocks and ensure system liveness. The scenario is a textbook case of ``lock inversion'', a term often used to describe such deadlock-prone patterns in concurrent programming.

\item[Pseudo Code:]{\phantom{.}}

\centering
\begin{minipage}{0.350\textwidth}
\begin{algorithm}[H]
\caption*{Thread 0}
\begin{algorithmic}[1]
\State Monitor.Enter(mutex1);
\State Monitor.Enter(mutex2);
\State critical\_section();
\State Monitor.Exit(mutex1);
\State Monitor.Exit(mutex2);
\end{algorithmic}
\end{algorithm}
\end{minipage}
\hspace{1cm}
\begin{minipage}{0.35\textwidth}
\begin{algorithm}[H]
\caption*{Thread 1}
\begin{algorithmic}[1]
\State Monitor.Enter(mutex2);
\State Monitor.Enter(mutex1);
\State critical\_section();
\State Monitor.Exit(mutex2);
\State Monitor.Exit(mutex1);
\end{algorithmic}
\end{algorithm}
\end{minipage}

\end{description}

\subsection{Lost Signal: Deadlock from Missed Signal in Condition Variable Coordination}

\label{sec:prob12}

\begin{description}
\item[Description:] Two threads coordinate using a shared condition variable. Thread 0 waits for a flag to become true using an \texttt{if} statement and then calls \texttt{wait()}. Thread 1 sets the flag and sends a notification using \texttt{notify\_all()}. If Thread 1 sends the signal before Thread 0 begins waiting, the signal is lost, and Thread 0 waits indefinitely.

\item[Effect:] A deadlock, as Thread 0 never receives the signal it depends on.

\item[Root Cause:] A weak guard: Thread 0 fails to re-check the condition after waking and uses an \texttt{if} statement instead of a \texttt{while} loop to guard the wait.

\item[Insight:] This problem reinforces a key principle in concurrent programming: condition variables must be used with guarded waits that revalidate the condition upon waking. This ensures correctness even in the presence of spurious wakeups or early notifications.

\item[Pseudo Code:]{\phantom{.}}

\begin{minipage}{0.48\textwidth}
\begin{algorithm}[H]
\caption*{Thread 0 (Waiter - Weak Guard)}
\begin{algorithmic}[1]
\State lock(mutex)
\If{flag == false}
\State cv.wait(mutex) \Comment{Bug: only checks once}
\EndIf
\State proceed\_assuming\_flag\_true()
\State unlock(mutex)
\end{algorithmic}
\end{algorithm}
\end{minipage}
\hspace{0.5cm}
\begin{minipage}{0.35\textwidth}
\begin{algorithm}[H]
\caption*{Thread 1 (Signaler)}
\begin{algorithmic}[1]
\State lock(mutex)
\State flag $\gets$ true
\State cv.notify\_all()
\State unlock(mutex)
\end{algorithmic}
\end{algorithm}
\end{minipage}
\end{description}

\subsection{Partial Lock: Race Condition from Insufficient Lock Coverage}

\label{sec:prob4}

\begin{description}
\item[Description:] Two threads manipulate a shared variable \texttt{i} under a locking mechanism. Thread 0 increments \texttt{i} by 2 and checks whether \texttt{i == 5}, while Thread 1 decrements \texttt{i} by 1. Although both threads use a lock, the locking does not encompass all relevant operations or ensure proper coordination between them. As a result, the interleaving of operations can lead to unexpected values of \texttt{i}, potentially triggering assertion failures.

\item[Effect:] An unexpected data or incorrect computation, as the shared state evolves in ways not anticipated by the program logic.

\item[Root Cause:] A missing or weak guard due to the lock is not applied consistently across all accesses and updates to the shared variable, allowing unsafe interleaving.

\item[Insight:] This example illustrates that merely using locks is not enough; they must be applied comprehensively and consistently to protect all shared state interactions.

\item[Pseudo Code:]{\phantom{.}}

\begin{minipage}{0.60\textwidth}
\begin{algorithm}[H]
\caption*{Thread 0}
\begin{algorithmic}[1]
\While{true}
  \State Monitor.Enter(mutex)
  \State $i \gets i + 2$
  \State critical\_section()
  \If{$i = 5$}
    \State Debug.Assert(false) \Comment{BUG: This assert can fail}
  \EndIf
  \State Monitor.Exit(mutex)
\EndWhile
\end{algorithmic}
\end{algorithm}
\end{minipage}
\hspace{0.25cm}
\begin{minipage}{0.30\textwidth}
\begin{algorithm}[H]
\caption*{Thread 1}
\begin{algorithmic}[1]
\While{true}
  \State Monitor.Enter(mutex)
  \State $i \gets i - 1$
  \State critical\_section()
  \State Monitor.Exit(mutex)
\EndWhile
\end{algorithmic}
\end{algorithm}
\end{minipage}
\end{description}

\subsection{Phantom Permit: Mutual Exclusion Violation from Semaphore Misuse}

\label{sec:prob16}

\begin{description}

\item[Description:] Two threads share a binary semaphore intended to serialise entry to a critical section. Thread~0 performs the canonical \texttt{Wait}–critical section–\texttt{Release} sequence, preserving mutual exclusion. Thread~1, by contrast, invokes \texttt{Wait(timeout)}. If the timeout expires, it nevertheless executes \texttt{Release}, effectively inserting an extra permit into the semaphore (a “phantom’’ permit).

\item[Effect:] Concurrent access arises when the phantom permit allows both threads to enter the critical section simultaneously, enabling interleaved operations that can corrupt shared state or violate higher-level invariants.

\item[Root Cause:] The defect is rooted in a misuse of concurrency primitives: issuing \texttt{Release} without first holding the semaphore breaks the required one-to-one pairing of \texttt{Wait}/\texttt{Release}. This increases the semaphore's count spuriously and defeats its mutual-exclusion guarantee.

\item[Insight:] Correct semaphore protocols demand that every \texttt{Release} correspond to a successful \texttt{Wait}. Introducing time-limited waits without compensating logic must be done carefully; otherwise, phantom permits can emerge and silently undermine critical-section protection.

\item[Pseudo Code:]{\phantom{.}}

\begin{minipage}{0.35\textwidth}
\begin{algorithm}[H]
\caption*{Thread 0 (Acquirer)}
\begin{algorithmic}[1]
\While{semaphore == 0}
\State wait()
\EndWhile
\State semaphore -= 1
\State critical\_section()
\State semaphore += 1
\end{algorithmic}
\end{algorithm}
\end{minipage}
\hspace{0.5cm}
\begin{minipage}{0.50\textwidth}
\begin{algorithm}[H]
\caption*{Thread 1 (Timed Failer)}
\begin{algorithmic}[1]
\If{timeout}
\State \textbf{/* never acquired semaphore */}
\State semaphore += 1 \Comment{BUG: false release}
\EndIf
\end{algorithmic}
\end{algorithm}
\end{minipage}
\end{description}

\subsection{Race-To-Wait: Deadlock from Non-Atomic Coordination}

\label{sec:prob13}

\begin{description}
\item[Description:] Two threads attempt to synchronize based on a shared counter \texttt{waiters}. Each thread increments the counter and then waits for it to reach a specific value (e.g., 2) before proceeding. However, the increment and check operations on \texttt{waiters} are not atomic. Both threads may read the value 1 simultaneously before either has incremented it again, leading them both to wait forever for the counter to reach 2, which never happens.

\item[Effect:] A classic deadlock, even though no explicit locking mechanism is involved.

\item[Root Cause:] A non-atomic operation on shared state: the threads make decisions based on stale or incomplete views of shared memory.

\item[Insight:] This example highlights how even minimalistic, lock-free coordination can result in liveness failures if atomicity is not respected.

\item[Pseudo Code:]{\phantom{.}}

\begin{minipage}{0.45\textwidth}
\begin{algorithm}[H]
\caption*{Thread 0}
\begin{algorithmic}[1]
\State temp $\gets$ waiters
\State waiters $\gets$ temp + 1 \Comment{BUG: non-atomic}
\If{waiters < 2}
\State wait()
\EndIf
\end{algorithmic}
\end{algorithm}
\end{minipage}
\hspace{0.5cm}
\begin{minipage}{0.40\textwidth}
\begin{algorithm}[H]
\caption*{Thread 1}
\begin{algorithmic}[1]
\State temp $\gets$ waiters
\State waiters $\gets$ temp + 1 \Comment{Same bug}
\If{waiters < 2}
\State wait()
\EndIf
\end{algorithmic}
\end{algorithm}
\end{minipage}
\end{description}

\subsection{Racy Increment: Race Condition from Non-Atomic Compound Operations}

\label{sec:prob2}

\begin{description}
\item [Description:]This problem illustrates a subtle but critical flaw in assuming that compound operations are atomic. Two threads execute the expression \texttt{a = a + 1; if (a == 1) enter critical section}, intending to allow only the first thread that increments \texttt{a} to 1 to enter the critical section. However, this logic fails under concurrent execution because the operation \texttt{a = a + 1} is not atomic-it decomposes into a sequence of read, increment, and write steps. If both threads interleave during these steps, they may each observe \texttt{a} as 0, increment it to 1, and both proceed into the critical section.

\item[Effect:] A concurrent access, where both threads enter a region that was intended to be accessed by only one. This leads to unexpected data, as the shared state is manipulated under the false assumption of exclusivity.

\item[Root Cause:] A non-atomic operation stemming from the non-atomicity of the increment-and-check sequence. Without synchronization, the interleaving of operations allows both threads to satisfy the condition \texttt{a == 1} simultaneously.

\item[Insight:] This example underscores the importance of using atomic operations or explicit synchronization mechanisms, such as locks or atomic primitives, when accessing shared variables. It also highlights how deceptively simple code can harbor concurrency bugs if the underlying memory operations are not properly understood.

\item[Pseudo Code:]{\phantom{.}}

\centering
\begin{minipage}{0.55\textwidth}
\begin{algorithm}[H]
\caption*{Thread 0}
\begin{algorithmic}[1]
\State temp $\gets a$
\State temp $\gets$ temp $+ 1$
\State $a \gets$ temp\hspace{0.25em} \(\triangleright\)BUG: non-atomic update may interleave

\If{$a = 1$}
    \State critical\_section()
\EndIf
\end{algorithmic}
\end{algorithm}
\end{minipage}
\hspace{0.5cm}
\begin{minipage}{0.35\textwidth}
\begin{algorithm}[H]
\caption*{Thread 1 (Expanded Assignment)}
\begin{algorithmic}[1]
\State temp $\gets a$
\State temp $\gets$ temp $+ 1$
\State $a \gets$ temp \Comment{BUG:  same}
\If{$a = 1$}
    \State critical\_section()
\EndIf
\end{algorithmic}
\end{algorithm}
\end{minipage}
\label{fig:non-atomic-update}
\end{description}

\subsection{Semaphore Leak: Mutual Exclusion Violation from Semaphore Misuse}

\label{sec:prob7}

\begin{description}
\item[Description:] Involves two threads using a semaphore to control access to a critical section. Thread 0 follows the standard \texttt{Wait}–critical section–\texttt{Release} pattern. Thread 1, however, performs a time-limited \texttt{Wait} and calls \texttt{Release} regardless of whether it successfully acquired the semaphore.

\item[Effect:] This behavior can corrupt the semaphore's internal count, allowing multiple threads to enter the critical section simultaneously, a clear concurrent access.

\item[Root Cause:] A misuse of primitive: releasing a semaphore without a corresponding acquisition violates the expected one-to-one pairing of \texttt{Wait} and \texttt{Release}.

\item[Insight:] This example underscores the importance of maintaining strict discipline when using semaphores. Any deviation from the expected protocol can compromise the integrity of the synchronization mechanism.

\item[Pseudo Code:]{\phantom{.}}

\begin{minipage}{0.30\textwidth}
\begin{algorithm}[H]
\caption*{Thread 0}
\begin{algorithmic}[1]
\While{true}
  \State semaphore.Wait()
  \State critical\_section()
  \State semaphore.Release()
\EndWhile
\end{algorithmic}
\end{algorithm}
\end{minipage}
\hspace{0.5cm}
\begin{minipage}{0.60\textwidth}
\begin{algorithm}[H]
\caption*{Thread 1}
\begin{algorithmic}[1]
\While{true}
  \If{semaphore.Wait(500)} \Comment{Wait with timeout}
    \State critical\_section()
    \State semaphore.Release()
  \Else
    \State semaphore.Release() \Comment{BUG: release without own}
  \EndIf
\EndWhile
\end{algorithmic}
\end{algorithm}
\end{minipage}
\end{description}

\subsection{Shared Counter: Mutual Exclusion Violation from Unsynchronized Counter}

\label{sec:prob9}

\begin{description}
\item[Description:] Involves two threads incrementing a shared counter and entering a critical section based on different thresholds, one at a count of 5, the other at 3. The counter is not protected by any synchronization mechanism, allowing updates to interleave unpredictably.

\item[Effect:] Both threads may enter the critical section simultaneously or at unintended times, leading to a concurrent access and unexpected data.

\item[Root Cause:]A race condition due to the non-atomic operation and check of the shared counter.

\item[Insight:] This example demonstrates the necessity of synchronizing access to shared counters, especially when control flow decisions depend on their values. Without atomicity, even simple arithmetic can lead to concurrency failures.

\item[Pseudo Code:]{\phantom{.}}

\begin{minipage}{0.45\textwidth}
\begin{algorithm}[H]
\caption*{Five-Headed Dragon}
\begin{algorithmic}[1]
\While{true}
  \State counter $\gets$ counter $+ 1$
  \If{counter == 5}
    \State critical\_section()
  \EndIf
\EndWhile
\end{algorithmic}
\end{algorithm}
\end{minipage}
\hspace{0.5cm}
\begin{minipage}{0.35\textwidth}
\begin{algorithm}[H]
\caption*{Three-Headed Dragon}
\begin{algorithmic}[1]
\While{true}
  \State counter $\gets$ counter $+ 1$
  \If{counter == 3}
    \State critical\_section()
  \EndIf
\EndWhile
\end{algorithmic}
\end{algorithm}
\end{minipage}
\end{description}

\subsection{Shared Flag: Mutual Exclusion Violation from Weak Boolean Flag Guard}

\label{sec:prob3}

\begin{description}
\item[Description:] Demonstrates the inadequacy of using a simple Boolean flag to enforce mutual exclusion. Two threads share a flag and use it to guard a critical section. Each thread spins while the flag is \texttt{true}, sets it to \texttt{true}, enters the critical section, and then resets it to \texttt{false}. However, the check (\texttt{flag != false}) and the update (\texttt{flag = true}) are not atomic. If one thread is preempted after checking the flag but before setting it, the other thread may also pass the check and set the flag, resulting in both threads entering the critical section concurrently.

\item[Effect:] A concurrent access, where the critical section is accessed simultaneously by multiple threads, leading to potential data corruption or logic errors.

\item[Root Cause:] A weak guard-the synchronization mechanism fails to ensure atomicity between the check and the update. This highlights the need for atomic test-and-set operations or proper locking mechanisms to enforce exclusive access.

\item[Insight:] This highlights the need for atomic test-and-set operations or proper locking mechanisms to enforce exclusive access.

\item[Pseudo Code:]{\phantom{.}}

\begin{minipage}{0.35\textwidth}
\begin{algorithm}[H]
\caption*{First Army}
\begin{algorithmic}[1]
\While{true}
  \While{flag $\neq$ false}
    \State/* busy wait */
  \EndWhile
  \State flag $\gets$ true
  \State critical\_section()
  \State flag $\gets$ false
\EndWhile
\end{algorithmic}
\end{algorithm}
\end{minipage}
\hspace{0.5cm}
\begin{minipage}{0.48\textwidth}
\begin{algorithm}[H]
\caption*{Second Army}
\begin{algorithmic}[1]
\While{true}
  \While{flag $\neq$ false}
    \State/* busy wait */
  \EndWhile
  \State flag $\gets$ true \Comment{BUG: both can pass check}
  \State critical\_section()
  \State flag $\gets$ false
\EndWhile
\end{algorithmic}
\end{algorithm}
\end{minipage}
\end{description}

\subsection{Signal-Then-Wait – Deadlock from Premature Signaling in Condition synchronization}

\label{sec:prob14}

\begin{description}
\item[Description:] Two threads coordinate using a shared flag and condition variable. The signaling thread sets the flag and calls \texttt{notify\_all()} before the waiting thread has entered the blocking wait. Although the waiting thread uses a correct \texttt{while} guard around the condition variable, the notification is missed entirely because the thread was not waiting yet.

\item[Effect:] A clear deadlock: the waiting thread blocks indefinitely, even though the condition it depends on was fulfilled. This occurs because condition variable signals do not persist - if a signal is sent before a thread is waiting, it is lost.

\item[Root Cause:] An incorrect ordering of commands - the signal is issued before the synchronization context is established. This leads to a fundamental timing mismatch between threads.

\item[Insight:] This pattern highlights that the timing of signal delivery in condition variable synchronization is critical. Signals must occur only after the corresponding wait condition has been armed, or the system risks falling into liveness failures such as deadlock.

\item[Pseudo Code:]{\phantom{.}}

\begin{minipage}{0.50\textwidth}
\begin{algorithm}[H]
\caption*{Thread 0 (Waiter)}
\begin{algorithmic}[1]
\State lock(mutex)
\While{flag == false}
    \State wait\_blocked $\gets$ true
    \State wait(cv, mutex) \Comment{BUG: signal already sent}
\EndWhile
\State use\_resource()
\State unlock(mutex)
\end{algorithmic}
\end{algorithm}
\end{minipage}
\hspace{0.5cm}

\begin{minipage}{0.65\textwidth}
\begin{algorithm}[H]
\caption*{Thread 1 (Signaler)}
\begin{algorithmic}[1]
\State flag $\gets$ true \Comment{BUG: condition updated before wait begins}
% \State yield()
\State lock(mutex)
\State notify\_all(cv)
\State unlock(mutex)
\end{algorithmic}
\end{algorithm}
\end{minipage}
\end{description}

\subsection{Sleeping Guard: Deadlock from Missing or Weak Guard}

\label{sec:prob15}

\begin{description}
\item[Description:] Presents a subtle but powerful failure in the use of condition synchronization. A consumer thread checks a queue, and if it's empty, sets a \texttt{waiting} flag and waits. A producer thread checks for the flag and enqueues data. The issue occurs if the producer enqueues a new item before the consumer sets the flag; the consumer misses the notification and remains blocked indefinitely.

\item[Effect:] A classic deadlock, in which the consumer thread remains permanently blocked waiting for a signal that was sent before it armed the condition, while the producer continues indefinitely, leaving the system with no forward progress.

\item[Root Cause:] A missing or weak guard: the consumer waits based solely on a flag without rechecking the real shared resource (the queue). In such designs, the wait must be governed by a guard that accurately reflects the synchronization invariant, and it must be re-evaluated after any wake-up event.

\item[Insight:] Without such a recheck, typically enforced with a \texttt{while} loop, the thread risks sleeping forever, even though the condition it depends on has already been satisfied.

\item[Pseudo Code:]{\phantom{.}}

\begin{minipage}{0.52\textwidth}
\begin{algorithm}[H]
\caption*{Consumer}
\begin{algorithmic}[1]
\If{queue.empty()}
\State waiting $\gets$ true
\State sleep() \Comment{BUG: doesn't recheck Q}
\EndIf
\State item $\gets$ queue.pop()
\State process(item)
\end{algorithmic}
\end{algorithm}
\end{minipage}
\hspace{0.5cm}
\begin{minipage}{0.35\textwidth}
\begin{algorithm}[H]
\caption*{Producer}
\begin{algorithmic}[1]
\State queue.push(item)
\If{waiting}
\State waiting $\gets$ false
\EndIf
\end{algorithmic}
\end{algorithm}
\end{minipage}
\end{description}

\section{Limitations of the Proposed Approach}

While our approach to bug amplification demonstrates strong empirical performance across a diverse set of concurrency problems, it is important to acknowledge its current limitations and boundaries of applicability.

\textbf{Dependence on Parameter Sensitivity.} Our method assumes that the probability of bug manifestation is meaningfully influenced by the input parameters exposed to the test generation engine. For systems where concurrency faults are insensitive to external parameters (e.g., bugs that manifest only due to internal scheduler decisions or deep state interactions), our black-box approach may offer limited leverage.

\textbf{Curse of Dimensionality.} As the dimensionality of the input space increases, learning an accurate regression model under a fixed testing budget becomes increasingly difficult. While our ensemble classifier demonstrated strong generalization in the studied benchmarks, its performance may degrade in higher-dimensional or sparsely populated input spaces, particularly when failure-inducing regions are extremely narrow.

\textbf{Noise Sensitivity and Stochastic Feedback.} Although we mitigate stochasticity through repeated executions, our framework is still subject to noise in failure observations. In scenarios where bug triggering is both rare and erratic, the resulting label noise can impair the quality of the learned models. This sensitivity places limits on how well regression-based methods can capture the underlying failure structure, especially early in the learning process.

\textbf{Model Retraining Overhead.} The iterative nature of our learning-based method requires frequent retraining of the classifier during test generation. While not a bottleneck in our Python-based implementation, this could become a concern for large-scale systems or industrial deployments with tight performance constraints, especially when test executions are costly.

\textbf{No Schedule Control.} Unlike techniques such as systematic concurrency testing or randomized schedulers, our approach does not manipulate the thread scheduler or execution order. As a result, bugs that require specific interleavings to manifest may remain elusive unless those conditions can be indirectly induced through parameter variation.

Despite these limitations, our method provides a practical, non-invasive tool for increasing the likelihood of bug detection in concurrent systems. It complements existing techniques by offering a black-box, input-driven strategy that is easy to integrate and effective across a wide range of problem types.

\section{Summary and Conclusions}

This paper addresses a fundamental challenge in software testing: reliably detecting concurrency bugs that manifest under rare interleavings and elusive execution schedules. These failures, though often critical, are notoriously hard to reproduce. To tackle this, we propose a probabilistic reformulation of the test generation task, treating bug detection as a problem of searching for inputs with maximized failure probability. This shift enables both a principled evaluation of search heuristics and the design of more effective testing strategies.

To evaluate our approach, we introduced a carefully curated benchmark of 17 multithreaded programs, each exhibiting a different concurrency failure. These programs span diverse root causes and error types, and the benchmark was built to ensure broad coverage and realism. For each problem, we examined the effectiveness of four black-box test generation methods: brute-force ($BF$), genetic algorithm (\textit{GA}), simulated annealing (\textit{SA}), and an ensemble classifier (\textit{Ens}).Each method was executed in 50 independent trials per problem, producing a robust dataset for statistical comparison.

The central contribution of the paper lies in the design and implementation of the ensemble-guided test generation strategy. By treating bug-finding as a classification problem over test inputs, our method learns from past failures and adaptively focuses the search on high-potential areas of the input space. This method is fully black-box and does not require access to program internals. Our results demonstrate that this learning-based strategy consistently outperforms traditional heuristics across nearly all benchmark problems. Notably, \textit{Ens} achieves higher detection rates using fewer test executions and converges more quickly to effective test inputs.

We further introduced a set of four graph-based analysis techniques that offer a detailed view of the methods' behavior: per-problem success curves, comparisons of top-ranked test-cases, convergence dynamics, and statistical significance heatmaps. These visual tools enabled us to examine method effectiveness from multiple perspectives and to identify patterns in both algorithmic performance and problem hardness. The analysis reveals that \textit{Ens} not only provides early bug discovery but also maintains its advantage as the test budget increases, exhibiting both statistical robustness and practical scalability.

Finally, we propose a novel simulation-based search heuristic for continuous input spaces inspired by simulated annealing but guided by probabilistic failure gradients. This formulation opens avenues for future work in guided bug-amplification over high-dimensional input domains.

In conclusion, this paper contributes a new methodological framework for adaptive bug-amplification, introduces a reusable benchmark of concurrency problems, and provides compelling empirical evidence that ensemble-guided testing can substantially improve the reliability and efficiency of concurrency bug detection. We believe these findings advance the state-of-the-art in automated software testing and lay a foundation for broader adoption of machine-learning methods in fault localization and test generation.

%%%%%%%%%%%%%%%%%%%%%%%%%%%%%%%%%%%%%%%%%%

\section{Administrative Data}
\label{sec:methods}

%%%%%%%%%%%%%%%%%%%%%%%%%%%%%%%%%%%%%%%%%%
%% optional
%\supplementary{The following supporting information can be downloaded at:  \linksupplementary{s1}, Figure S1: title; Table S1: title; Video S1: title.}


% Only used for preprtints:
% \supplementary{The following supporting information can be downloaded at the website of this paper posted on \href{https://www.preprints.org/}{Preprints.org}.}

%\begin{tabularx}{\textwidth}{lll}
%\toprule
%\textbf{Name} & \textbf{Type} & \textbf{Description} \\
%\midrule
%S1 & Python script (.py) & Script of python source code used in XX \\
%S2 & Text (.txt) & Script of  code used to make Figure X \\
%S3 & Text (.txt) & Raw data from experiment X \\
%S4 & Video (.mp4) & Video demonstrating the hardware in use \\
%\bottomrule
%\end{tabularx}
%}

%%%%%%%%%%%%%%%%%%%%%%%%%%%%%%%%%%%%%%%%%%
\authorcontributions{Conceptualization, all = (Y.W., G.W., O.M., E.F., G.A. \& A.E.); methodology, all; software, Y.W.; validation, Y.W; formal analysis, all; investigation, Y.W.; resources, all; data curation, Y.W.; writing---original draft preparation, Y.W.; writing---review and editing, all; visualization, Y.W.; supervision, G.W.; project administration, all; funding acquisition, all. All authors have read and agreed to the published version of the manuscript.}

\funding{This research was partially funded by the Lynne and William Frankel Center for Computer Science and the Israeli Science Foundation grant No. 2714/19.}

\institutionalreview{Not applicable.}

\informedconsent{Not applicable.}

\dataavailability{We implemented a modular Python-based framework for conducting the amplification experiments
The original data presented in the study are openly available in \url{https://github.com/geraw/bug_amp}. 

The framework supports multiple amplification strategies, including Brute-Force ($BF$), Simulated Annealing (\textit{SA}), Genetic Algorithms (\textit{GA}), and an ensemble classifier-based method (\textit{Ens}). It is designed to be easily extensible, encouraging researchers and practitioners to contribute and experiment with their own search techniques. By following the provided templates and interface guidelines, users can seamlessly integrate new amplification strategies into the framework. We warmly invite the community to build upon our work and adapt the system to their specific needs.

All experiments were conducted on the high-performance computing (\texttt{}{HPC}) infrastructure at Ben-Gurion University of the Negev (BGU), which is managed using an internal \texttt{SLURM} workload manager. \texttt{SLURM} allowed us to schedule thousands of concurrent and independent test executions efficiently across a cluster of multi-core servers. The use of this environment significantly accelerated the experimentation process and allowed us to evaluate all methods consistently across all problem instances.

Although the experiments were originally executed in a \texttt{SLURM}-based cluster environment, the entire codebase is portable. It can be executed on any standard Linux or Windows machine or cloud platform using Python 3 and common scientific libraries, without the need for \texttt{SLURM}. The repository includes detailed instructions for reproducing the experiments, ensuring transparency and repeatability across different platforms.
}

\acknowledgments{During the preparation of this study, we used \texttt{ChatGPT} and \texttt{Gemini} for the purposes of searching for related work, coding, and writing. The authors have reviewed and edited the output and take full responsibility for the content of this publication.}

\conflictsofinterest{The authors declare no conflicts of interest.} 

%%%%%%%%%%%%%%%%%%%%%%%%%%%%%%%%%%%%%%%%%%
%% Optional

%% Only for journal Encyclopedia
%\entrylink{The Link to this entry published on the encyclopedia platform.}

\abbreviations{Abbreviations}{
The following abbreviations are used in this manuscript:
\\
\noindent
\begin{tabular}{@{}ll}
BF & Brute-Force \\
BGU & Ben-Gurion University of the Negev \\
CI & Continuous Integration \\
CLT &  Central Limit Theorem \\
Ens & Ensemble \\
GA & Genetic Algorithm \\
GP & Genetic Programming \\
LLN & Law of Large Numbers \\
HPC & High-Performance Computing \\
MLP & Multi-Layer Perceptron \\
PCT & Probabilistic Concurrency Testing \\
SA & Simulated Annealing \\
SD & Standard Deviation \\
SLURM & Simple Linux Utility for Resource Management \\
SMOTE & Minority Over-sampling Technique \\
% SOT & Sysntetic Oversampling Technique \\
SUT & System Under Test
\end{tabular}
}

%%%%%%%%%%%%%%%%%%%%%%%%%%%%%%%%%%%%%%%%%%
%% Optional
% \appendixtitles{no} % Leave argument "no" if all appendix headings stay EMPTY (then no dot is printed after "Appendix A"). If the appendix sections contain a heading then change the argument to "yes".
% \appendixstart
% \appendix
% \section[\appendixname~\thesection]{}
% \subsection[\appendixname~\thesubsection]{}
% The appendix is an optional section that can contain details and data supplemental to the main text---for example, explanations of experimental details that would disrupt the flow of the main text but nonetheless remain crucial to understanding and reproducing the research shown; figures of replicates for experiments of which representative data are shown in the main text can be added here if brief, or as Supplementary Data. Mathematical proofs of results not central to the paper can be added as an appendix.

% \section[\appendixname~\thesection]{}
% All appendix sections must be cited in the main text. In the appendices, Figures, Tables, etc. should be labeled, starting with ``A''---e.g., Figure A1, Figure A2, etc.

%%%%%%%%%%%%%%%%%%%%%%%%%%%%%%%%%%%%%%%%%%
%\isPreprints{}{% This command is only used for ``preprints''.
% \begin{adjustwidth}{-\extralength}{0cm}
\printendnotes[custom] % Un-comment to print a list of endnotes

\reftitle{References}


\bibliography{MDPIBugAmp} % filename of your .bib (no extension)

\begin{thebibliography}{55}
\bibitem{Gray1985}
Jim Gray. Why Do Computers Stop and What Can Be Done About It?. , 1985.


\bibitem{intermitent}
Bakhshi, Roozbeh and Kunche, Surya and Pecht, Michael. Intermittent Failures in Hardware and Software. Journal of Electronic Packaging, 2014.


\bibitem{Musuvathi2008}
Madanlal Musuvathi and Shaz Qadeer and Thomas Ball and Gerard Basler and Piramanayagam Arumuga Nainar and Iulian Neamtiu. Finding and Reproducing Heisenbugs in Concurrent Programs. 8th USENIX Symposium on Operating Systems Design and Implementation (OSDI 2008), 2008.


\bibitem{Burckhardt2010}
Sebastian Burckhardt and Pravesh Kothari and Madanlal Musuvathi and Santosh Nagarakatte. A Randomized Scheduler with Probabilistic Guarantees of Finding Bugs. 15th International Conference on Architectural Support for Programming Languages and Operating Systems (ASPLOS '10), 2010.


\bibitem{Godefroid2015}
Patrice Godefroid and Michael Y. Levin and David A. Molnar. Effective Testing for Concurrency Bugs. , 2015.


\bibitem{Elmas2013}
Tayfun Elmas and Jacob Burnim and George C. Necula and Koushik Sen. CONCURRIT: A Domain Specific Language for Reproducing Concurrency Bugs. 34th ACM SIGPLAN Conference on Programming Language Design and Implementation (PLDI '13), 2013.


\bibitem{Li2019DeepFL}
Xiaowen Li and Weihai Li and Yingjun Zhang and Lijie Zhang. DeepFL: Integrating Multiple Fault Diagnosis Dimensions for Deep Fault Localization. 28th ACM SIGSOFT International Symposium on Software Testing and Analysis (ISSTA '19), 2019.


\bibitem{Bottinger2018LearnFuzz}
Konstantin Böttinger and Patrice Godefroid and Rishabh Singh. Learn\&Fuzz: Machine Learning for Input Fuzzing. arXiv, 2018.


\bibitem{10.1145/2954679.2872374}
Tanakorn Leesatapornwongsa and Jeffrey F. Lukman and Shan Lu and Haryadi S. Gunawi. TaxDC: A Taxonomy of Non‑Deterministic Concurrency Bugs in Datacenter Distributed Systems. 51st ACM SIGPLAN Conference on Programming Language Design and Implementation (PLDI '16), 2016.


\bibitem{goldberg1989}
David E. Goldberg. Genetic Algorithms in Search, Optimization and Machine Learning. , 1989.


\bibitem{sipper2023eckity}
Sipper, Moshe and Green, Brian and Ronen, Yakir and Gat, Tomer and Hoffman, Shaked and Zohar, Noam. EC-KitY. SoftwareX, 2023.


\bibitem{Karafotias2015}
Giorgos Karafotias and Mark Hoogendoorn and A. E. Eiben. Parameter Control in Evolutionary Algorithms: Trends and Challenges. IEEE Transactions on Evolutionary Computation, 2015.


\bibitem{Elyasaf2023}
Elyasaf, Achiya and Farchi, Eitan and Margalit, Oded and Weiss, Gera and Weiss, Yeshayahu. Generalized Coverage Criteria for Combinatorial Sequence Testing. IEEE Transactions on Software Engineering, 2023.


\bibitem{Wasserstein2016}
Ronald L. Wasserstein and Nicole A. Lazar. The ASA's Statement on p‑Values: Context, Process, and Purpose. The American Statistician, 2016.


\bibitem{Liu2024TrickCatcher}
Kaibo Liu and Zhenpeng Chen and Yiyang Liu and Jie M. Zhang and Mark Harman and Yudong Han and Yun Ma and Yihong Dong and Ge Li and Gang Huang. LLM‑Powered Test Case Generation for Detecting Bugs in Plausible Programs. arXiv preprint arXiv:2404.10304, 2024.


\bibitem{Ouedraogo2025}
Wendkûuni C. Ouédraogo and Laura Plein and Kader Kaboré and Andrew Habib and Jacques Klein and David Lo and Tegawendé F. Bissyandé. Enriching Automatic Test Case Generation by Extracting Relevant Test Inputs from Bug Reports. Empirical Software Engineering, 2025.


\bibitem{Benavoli2015}
Alessio Benavoli and Giorgio Corani and Francesca Mangili. Should we really use post‑hoc tests based on mean‑ranks?. CoRR, 2015.


\bibitem{MightVanHorn2011}
Matthew Might and David Van Horn. A Family of Abstract Interpretations for Static Analysis of Concurrent Higher-Order Programs. Static Analysis (SAS 2011), 2011.


\bibitem{Bora2021LLVMHPC}
Utpal Bora and Shraiysh Vaishay and Saurabh Joshi and Ramakrishna Upadrasta. OpenMP Aware MHP Analysis for Improved Static Data‑Race Detection. 7th IEEE/ACM Workshop on the LLVM Compiler Infrastructure in HPC (LLVM‑HPC '21), 2021.


\bibitem{MatsakisKlock2014}
Nicholas D. Matsakis and Felix S. Klock II. The Rust Language. Ada Letters, 2014.


\bibitem{Tehrani2019DeepRace}
Ali Tehrani and Mohammed Khaleel and Reza Akbari and Ali Jannesari. DeepRace: Finding Data Race Bugs via Deep Learning. arXiv preprint arXiv:1907.07110, 2019.


\bibitem{Chen2020MUZZ}
Hongxu Chen and Shengjian Guo and Yinxing Xue and Yulei Sui and Cen Zhang and Yuekang Li and Haijun Wang and Yang Liu. MUZZ. 29th USENIX Security Symposium (USENIX Security '20), 2020.


\bibitem{Roemer2020SmartTrack}
Jake Roemer and Kaan Genç and Michael D. Bond. SmartTrack: Efficient Predictive Race Detection. 41st ACM SIGPLAN Conference on Programming Language Design and Implementation (PLDI '20), 2020.


\bibitem{OCallahan2017RR}
Robert O'Callahan and Chris Jones and Nathan Froyd and Kyle Huey and Albert Noll and Nimrod Partush. Engineering Record And Replay For Deployability. 2017 USENIX Annual Technical Conference (USENIX ATC '17), 2017.


\bibitem{Holzmann1997SPIN}
Gerard J. Holzmann. The Model Checker SPIN. IEEE Transactions on Software Engineering, 1997.


\bibitem{Clarke2001BMC}
Edmund M. Clarke and Armin Biere and Richard Raimi and Yunshan Zhu. Bounded Model Checking Using Satisfiability Solving. Formal Methods in System Design, 2001.


\bibitem{NamjoshiTrefler2016}
Kedar S. Namjoshi and Richard J. Trefler. Parameterized Compositional Model Checking. Tools and Algorithms for the Construction and Analysis of Systems (TACAS 2016), 2016.


\bibitem{Legay2019SMC}
Axel Legay and Anna Lukina and Louis‑Marie Traonouez and Junxing Yang and Scott A. Smolka and Radu Grosu. Statistical Model Checking. Computing and Software Science, 2019.


\bibitem{Xu2020KRACE}
Meng Xu and Sanidhya Kashyap and Hanqing Zhao and Taesoo Kim. KRACE: Data Race Fuzzing for Kernel File Systems. 2020 IEEE Symposium on Security and Privacy (SP), 2020.


\bibitem{Lu2008Learning}
Shan Lu and Soyeon Park and Eunsoo Seo and Yuanyuan Zhou. Learning from Mistakes: A Comprehensive Study on Real World Concurrency Bug Characteristics. 13th International Conference on Architectural Support for Programming Languages and Operating Systems (ASPLOS '08), 2008.


\bibitem{Musuvathi2008Heisenbugs}
Madanlal Musuvathi and Shaz Qadeer and Thomas Ball and Gerard Basler and Dirk R. Engler and Joseph C. Foster and Amit K. Ghosh. Finding and Reproducing Heisenbugs in Concurrent Programs. 8th USENIX Symposium on Operating Systems Design and Implementation (OSDI), 2008.


\bibitem{Tian2011RaceBench}
Yongjun Tian and Yongfei Yu and Peng Wang and Ruihong Zhou and Hui Jin and Tao Xie. RACEBENCH: A Benchmark Suite for Data Race Detection Tools. 19th ACM SIGSOFT Symposium and the 13th European Conference on Foundations of Software Engineering (ESEC/FSE '11), 2011.


\bibitem{Zhang2011ConSeq}
Wei Zhang and Chen Yao and Shan Lu and Jeff Huang and Tian Tan and Xu Liu. ConSeq: Detecting Concurrency Bugs Through Sequential Errors. 16th International Conference on Architectural Support for Programming Languages and Operating Systems (ASPLOS '11), 2011.


\bibitem{Lin2015JaConTeBe}
Ziyi Lin and Darko Marinov and Hao Zhong and Yuting Chen and Jianjun Zhao. JaConTeBe: A Benchmark Suite of Real-World Java Concurrency Bugs. 30th IEEE/ACM International Conference on Automated Software Engineering (ASE '15), 2015.


\bibitem{Just2014Defects4J}
Ren\'e. Defects4J. 2014 International Symposium on Software Testing and Analysis (ISSTA '14), 2014.


\bibitem{Madeiral2019Bears}
Fernanda Madeiral and Simon Urli and Marcelo de Almeida Maia and Martin Monperrus. BEARS: An Extensible Java Bug Benchmark for Automatic Program Repair Studies. 26th IEEE International Conference on Software Analysis, Evolution and Reengineering (SANER '19), 2019.


\bibitem{Karampatsis2020ManySStuBs4J}
Rafael‑Michael Karampatsis and Charles Sutton. How Often Do Single‑Statement Bugs Occur?: The ManySStuBs4J Dataset. 17th International Conference on Mining Software Repositories (MSR '20), 2020.


\bibitem{Tu2019GoConcurrency}
Tengfei Tu and Xiaoyu Liu and Linhai Song and Yiying Zhang. Understanding Real‑World Concurrency Bugs in Go. 24th International Conference on Architectural Support for Programming Languages and Operating Systems (ASPLOS '19), 2019.


\bibitem{Yuan2021GoBench}
Ting Yuan and Guangwei Li and Jie Lu and Chen Liu and Lian Li and Jingling Xue. GoBench: A Benchmark Suite of Real‑World Go Concurrency Bugs. 18th Annual IEEE/ACM International Symposium on Code Generation and Optimization (CGO '21), 2021.


\bibitem{Torres2021ConFuzzius}
Christof Ferreira Torres and Antonio Ken Iannillo and Arthur Gervais and Radu State. ConFuzzius: A Data Dependency-Aware Hybrid Fuzzer for Smart Contracts. 2021 IEEE European Symposium on Security and Privacy (EuroS\&P '21), 2021.


\bibitem{Sen2005DART}
Godefroid, Patrice and Klarlund, Nils and Sen, Koushik. DART: Directed Automated Random Testing. 2005 ACM SIGPLAN Conference on Programming Language Design and Implementation (PLDI), 2005.


\bibitem{Clarke2000Abstraction}
Clarke, Edmund M. and Grumberg, Orna and Jha, Somesh and Lu, Yuan and Veith, Helmut. Counterexample-Guided Abstraction Refinement. 12th International Conference on Computer Aided Verification (CAV), 2000.


\bibitem{Bianchi2017ConCrashSearch}
Francesco A. Bianchi and Mauro Pezz\`e. A Search-Based Approach to Reproduce Crashes in Concurrent Programs. 11th Joint Meeting on Foundations of Software Engineering (ESEC/FSE), 2017.


\bibitem{Amalfitano2023}
Amalfitano, Domenico and Faralli, Stefano and Hauck, Jean Carlo Rossa and Matalonga, Santiago and Distante, Damiano. Artificial Intelligence Applied to Software Testing: A Tertiary Study. ACM Computing Surveys, 2023.


\bibitem{Lee2022}
Lee, Sanghoon and Zhang, Hui and Viswanathan, Mahesh. Probabilistic Concurrency Testing for Weak Memory Programs. 28th ACM SIGPLAN Symposium on Principles and Practice of Parallel Programming (PPoPP), 2023.


\bibitem{Chen2023}
Chen, Yuzhe and Liu, Shuhan and Gan, Qiao. Effective Concurrency Testing for Go via Directional Primitive Scheduling. 38th IEEE/ACM International Conference on Automated Software Engineering (ASE), 2023.


\bibitem{Kumar2025}
Kumar, Ravi and Lee, Jungho and Padhye, Rohan. Fray: An Efficient General-Purpose Concurrency Testing Platform for JVM. arXiv, 2025.


\bibitem{Xu2025}
Xu, Jinyang and Wolff, David and Han, Xinyu and Li, Jie and Roychoudhury, Abhik. Concurrency Testing in the Linux Kernel via eBPF. arXiv, 2025.


\bibitem{Han2024}
Han, Tianshuo and Gong, Xiangyu and Liu, Jie. CARDSHARK: Understanding and Stabilizing Linux Kernel Concurrency Bugs Against the Odds.  33rd USENIX Security Symposium (USENIX Security 24), 2024.


\bibitem{ramesh2025unveiling}
Ramesh, Arjun and Huang, Tianshu and Riar, Jaspreet and Titzer, Ben L. and Rowe, Anthony. Unveiling Heisenbugs with Diversified Execution. ACM on Programming Languages, 2025.


\bibitem{Rasheed2023}
Rasheed, Shawn and Dietrich, Jens and Tahir, Amjed. On the Effect of Instrumentation on Test Flakiness. 2023 IEEE/ACM International Conference on Automation of Software Test (AST), 2023.


\bibitem{Shashidhar2021}
Shashank, Sai Shashidhar and Sachdeva, Jatin and Mukherjee, Suvam and Deligiannis, Pantazis. Nekara: A Generalized Concurrency Testing Library. 36th IEEE/ACM International Conference on Automated Software Engineering (ASE), 2021.


\bibitem{rare-event-simulation}
Heidelberger, Philip. Fast simulation of rare events in queueing and reliability models. ACM Trans. Model. Comput. Simul., 1995.


\bibitem{statistical-model-checking}
Håkan L.S. Younes and Reid G. Simmons. Statistical probabilistic model checking with a focus on time-bounded properties. Information and Computation, 2006.


\bibitem{zhao2025selectively}
Zhao, Huan and Wolff, Dylan and Mathur, Umang and Roychoudhury, Abhik. Selectively Uniform Concurrency Testing. Proceedings of the ACM on Programming Languages (ASPLOS), 2025.

\end{thebibliography}
%=====================================
% \bibliography{MDPIBugAmp.bib}

% If authors have biography, please use the format below
%\section*{Short Biography of Authors}
%\bio
%{\raisebox{-0.35cm}{\includegraphics[width=3.5cm,height=5.3cm,clip,keepaspectratio]{Definitions/author1.pdf}}}
%{\textbf{Firstname Lastname} Biography of first author}
%
%\bio
%{\raisebox{-0.35cm}{\includegraphics[width=3.5cm,height=5.3cm,clip,keepaspectratio]{Definitions/author2.jpg}}}
%{\textbf{Firstname Lastname} Biography of second author}

% For the MDPI journals use author-date citation, please follow the formatting guidelines on http://www.mdpi.com/authors/references
% To cite two works by the same author: \citeauthor{ref-journal-1a} (\citeyear{ref-journal-1a}, \citeyear{ref-journal-1b}). This produces: Whittaker (1967, 1975)
% To cite two works by the same author with specific pages: \citeauthor{ref-journal-3a} (\citeyear{ref-journal-3a}, p. 328; \citeyear{ref-journal-3b}, p.475). This produces: Wong (1999, p. 328; 2000, p. 475)

%%%%%%%%%%%%%%%%%%%%%%%%%%%%%%%%%%%%%%%%%%
%% for journal Sci
%\reviewreports{\\
%Reviewer 1 comments and authors' response\\
%Reviewer 2 comments and authors' response\\
%Reviewer 3 comments and authors' response
%}
%%%%%%%%%%%%%%%%%%%%%%%%%%%%%%%%%%%%%%%%%%
\PublishersNote{}
%\isPreprints{}{% This command is only used for ``preprints''.
% \end{adjustwidth}
\end{document}